\documentclass[a4paper]{article}
\usepackage[english]{babel}
\usepackage{natbib}
\usepackage{url}
\usepackage{graphicx}
\usepackage{amsmath}
\begin{document}

\title{Single precision arithmetic in ECHAM radiation reduces runtime and energy consumption}

\author{Alessandro Cotronei, Thomas Slawig (ts@informatik.uni-kiel.de),\\
Kiel Marine Science (KMS) -- Centre for Interdisciplinary Marine Science,\\
Dep. of Computer Science, Kiel University, 24098 Kiel, Germany.}

\maketitle

\begin{abstract}
We converted the radiation part of the atmospheric model 
ECHAM to single precision arithmetic. We analyzed different conversion strategies and finally used a step by step change of all modules, subroutines and functions. We found out that a small code portion still requires higher precision arithmetic. We generated code that can be easily changed from double to single precision and vice versa, basically using a simple switch in one module. We compared the output of the single precision version in the coarse resolution with observational data and with the original double precision code. The results of both versions are comparable. We extensively tested different parallelization options  with respect to the possible performance gain, in both coarse and low resolution. The single precision radiation itself was accelerated by about 40\%, whereas the speed-up for the whole ECHAM model using the converted radiation achieved 18\% in the best configuration. We further measured the energy consumption, which could also be reduced.
\end{abstract}

\section{Introduction}  

\begin{figure}[t]
\includegraphics[width=8.3cm]{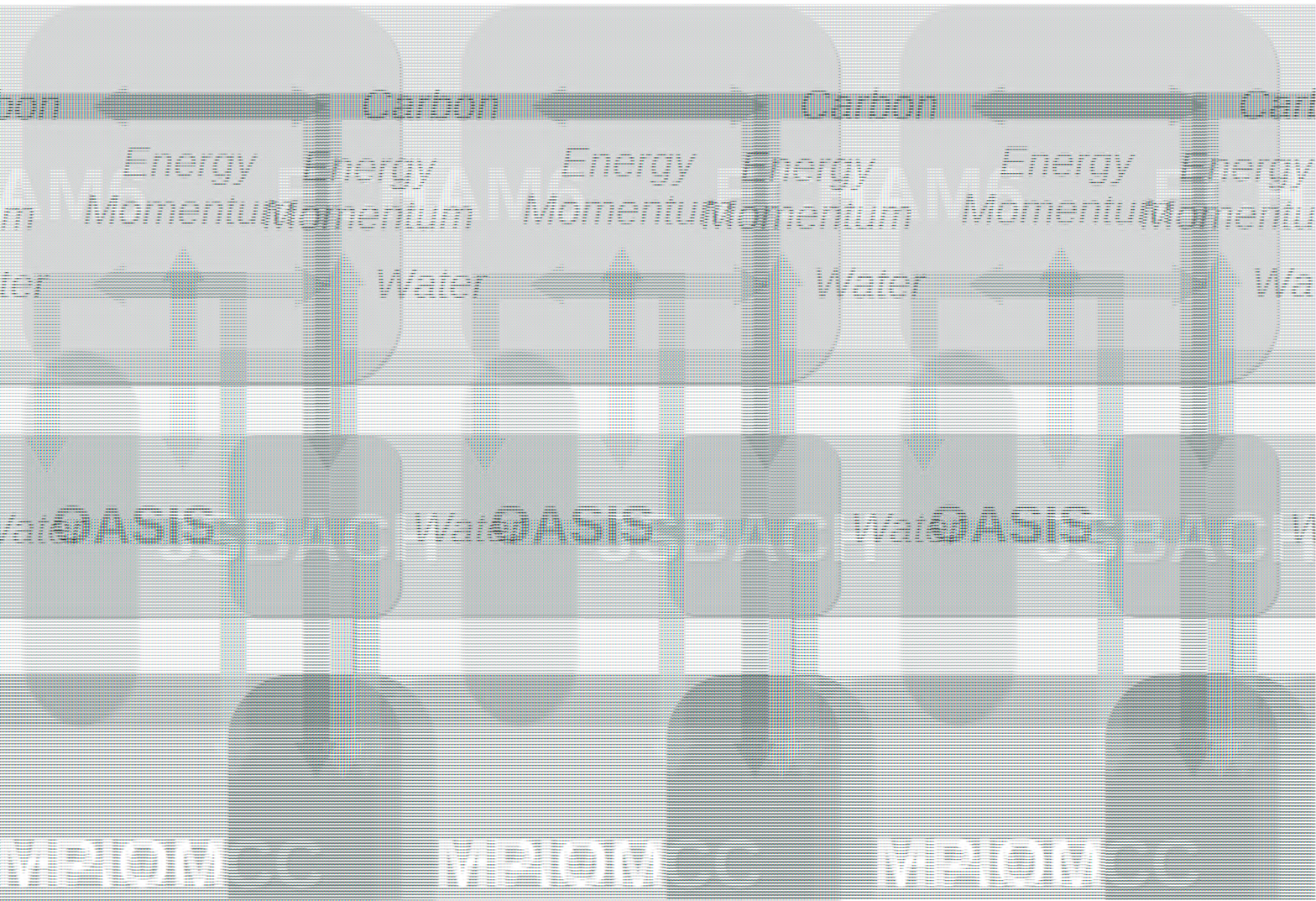}
\caption{Schematic of the structure of the Earth System Model MPI-ESM with atmospheric component ECHAM, terrestrial vegetation model JSBACH, ocean model MPI-OM, marine biogeochemical model HAMOCC, and OASIS coupler.}
   \label{fig:MPI-ESM}
\end{figure}

The atmospheric model ECHAM 
was developed at the Max Planck Institute for Meteorology (MPI-M) in Hamburg.
Its development started in 1987 as a branch of a global weather forecast model of the European Centre for Medium-Range Weather Forecasts (ECMWF), thus leading to the acronym (EC for ECMWF, HAM for Hamburg).  The model is  used in different Earth System Models (ESMs) as atmospheric component, e.g., in the MPI-ESM also developed at the MPI-M, see Figure \ref{fig:MPI-ESM}. The current version is ECHAM 6 
\citep{Stevens2013}. For  a detailed list on ECHAM publications we refer to the homepage of the institute (\url{mpimet.mpg.de}). Version 5 of the model was used in the 4th Assessment Report of the Intergovernmental Panel on Climate Change \citep{IPCC2007}, version 6 in the Coupled Model Intercomparison Project CMIP \citep{CMIP2019}.

Motivation for our work was the usage of ECHAM for long-time paleo-climate simulations in the German national climate modeling initiative 
"PalMod: From the Last Interglacial to the Anthropocene -- Modeling a Complete Glacial Cycle" (\url{www.palmod.de}). 
The aim of this initiative is to perform simulations for a complete glacial cycle, i.e. about 120'000 years, with fully coupled ESMs.

The feasibility of  long-time simulation runs highly depends on the computational performance of the used models.
As a consequence, one main focus in the PalMod project  is to decrease the runtime of the model components and the coupled ESMs.

\begin{figure}[t]
\includegraphics[width=8.3cm]{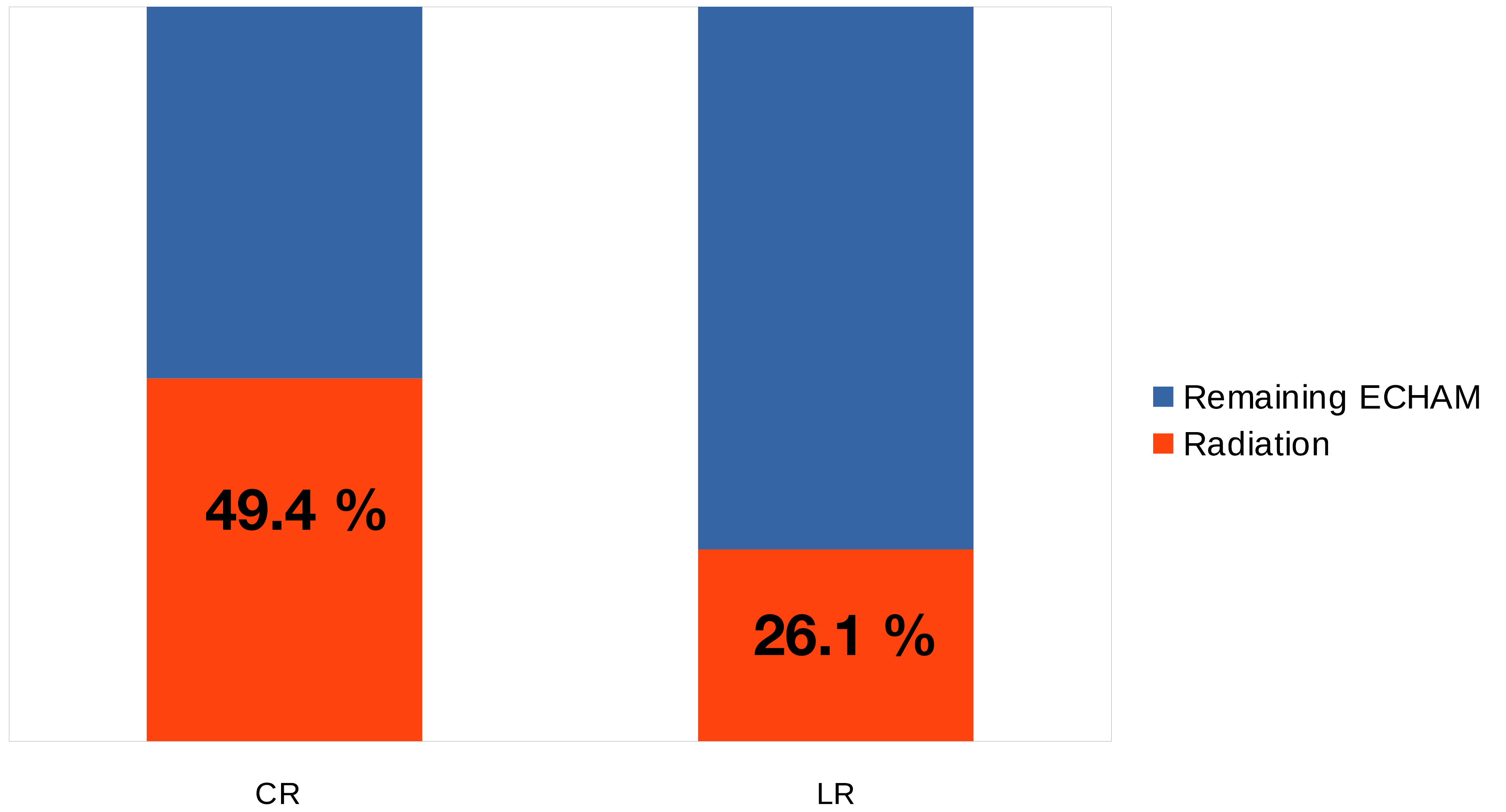}
\caption{Time consumption of the radiation part w.r.t. the whole ECHAM model  in coarse (CR) and low resolution (LR) of standard PalMod experiments.
The difference occurs since in these configuration the radiation part is called every 2 hours, i.e., only every fourth (in CR) or  every eighth time-step (LR).}
   \label{fig:radiation-time}
\end{figure}

In ESMs that use ECHAM, the part of the computational time that is used by the latter is significant. It can be close to 75\% in some configurations.
Within ECHAM itself, the radiation takes the most important part of the computational time.  As a consequence, the  radiation part is not called in every  time-step  in the current ECHAM setting. Still, its part of the overall ECHAM runtime is relevant, see Figure \ref{fig:radiation-time}.

In the PalMod project,  two different strategies to improve the performance of the radiation part are investigated: One is to 
run the radiation in parallel on different processors, the other one is the conversion to single precision arithmetic we present in this paper.
For both purposes, the radiation code was isolated from the rest of the ECHAM model. This technical procedure is not described here.

The motivation for the idea to improve the computational performance of ECHAM by a conversion to reduced arithmetic precision was the work of \citet{Vana2017}. In this paper, the authors report on the conversion of the Integrated Forecasting System (IFS)  model   to single precision, observing a runtime reduction of 40\% in short-time runs of 12 or 13 months, and a good  match of the output with observational data. Here, the terms \emph{single} and \emph{double precision} refer to the IEEE-754 standard for floating point numbers \citep{ieee2019}, therein also named  \emph{binary64} and \emph{binary32}, respectively. In the IEEE standard, also an even more reduced precision, the \emph{half precision (binary16)} format is defined. 
 The IFS model, developed also by the ECMWF, is comparable to ECHAM in some respect since it also uses a combination of spectral and grid-point-based discretization. A similar performance gain of 40\% was reported by   \citet{Ruedisuehli2014} with the COSMO model that is used by the German and Swiss weather forecast services. The authors also validated the model output by comparing it to observations and the original model version. 
 
 Recently, the usage of reduced precision arithmetic has gained interest for a variety of reasons. Besides the effect on the runtime, also a reduction of  energy consumption is mentioned, see, e.g.,  \citet{Fagan2016}, who reported a reduction by  about 60\%. 
 In the growing field of machine learning,  single or even more reduced precision is used to save both computational effort as well as memory, motivated by the  usage of Graphical processing Units (GPUs). \citet{Dawson2017} used reduced precision   to evaluate model output uncertainty. For this purpose, the authors developed a software where a variable precision is available, but a  positive effect on the model runtime was not their concern. 

The process of porting a  simulation code to a different precision highly depends on the design of the code and the way how basic principles of software engineering have been  followed during the implementation process. These are modularity, use of clear subroutine interfaces, way of data transfer via parameter list or global variables etc.
The main problem in legacy codes with a long history (as ECHAM) is that these principles usually were not applied very strictly. This is a general problem in computational science software, not only in climate modeling, see for example  \citet{Johanson2018}.

Besides the desired  performance gain, a main criterion to assess the result of the conversion to reduced precision is the validation of the results, i.e., their  differences to observational data and the output of the original, double precicion version. We  carried out  experiments on short time scales of 30 years with a 10 years spin-up. 
It has to be taken into account that after the conversion, a model tuning process (in fully coupled version) might be necessary. This will require a significant amount of work to obtain an ESM that produces a reasonable climate, see, e.g., \cite{Mauritsen2012} for a description of the tuning of the MPI-ESM.

The structure of the paper is as follows:
In the following section, we describe the situation from where our study and conversion started. In Section \ref{sec:strategies}, we give an overview about possible strategies to perform a conversion to single precision, discuss their applicability, and finally the motivation for the direction we took.  
In Section  \ref{sec:changes}, we describe   changes  that were necessary at some parts of the code due to certain used constructs or libraries, and in Section \ref{sec:higherprecision}  the parts of the code that need to remain in higher precision. In Section \ref{sec:results}, we present the obtained results w.r.t. performance gain, output validation and energy consumption.
At the end of the paper, we summarize our work and draw some conclusions.

\section{Used configuration of ECHAM}
The current major version of ECHAM, version  6, is described in \cite{Stevens2013}. ECHAM is a combination of a spectral  and a  grid-based finite difference model. It can be  used in five resolutions, ranging from the \emph{coarse resolution} (CR) or T31 (i.e., a truncation to 31 wave numbers in  the spectral part, corresponding to a horizontal spatial resolution of 96 $\cdot$ 48 points in longitude and latitude) up to XR or T255. We present  results for the CR and  LR (\emph{low resolution}, T63, corresponding to 192 $\cdot$ 96 points) versions.  Both   use 47 vertical layers and (in our setting)  time steps of 30 and 15 min., respectively. 

ECHAM6 is written in free format Fortran and conforms to the Fortran 95 standard \citep{metcalf2018}.
It consists of about 240'000  lines of code (including approximately 71'000 lines of the JSBACH vegetation model) and uses 
a number  of external libraries including LAPACK, BLAS (for linear algebra), MPI (for parallelization), and NetCDF (for in- and output). The radiation part that we converted contains about 30'000 lines of code and uses external libraries as well.

The basis ECHAM version we used is derived from the stand-alone version ECHAM-6.03.04p1. In this basis version, the radiation was modularly separated from the rest of ECHAM. This  offers the option to run the radiation and 
the remaining part of the model on different processors in order to reduce the running time by parallelization, but
 also maintains the possibility of running the ECHAM components sequentially. It was shown that the sequential version reproduces bit-identical results with the original code.

All the results presented below are evaluated with the Intel Fortran compiler 18.0 (update 4) \citep{Intel2017} on the supercomputer HLRE-3 \emph{Mistral}, located at the German Climate Computing Center (DKRZ), Hamburg.
All experiments used the so-called  \emph{compute} nodes of the machine.
\section{Strategies for conversion to single precision}
\label{sec:strategies}
In this section we give an overview of possible strategies  for the conversion of a simulation code (as the radiation part of ECHAM) to single precision arithmetic. We describe  the problems that occurred while applying them to the ECHAM radiation part.  At the end, we describe the strategy that finally turned out to be successful. 
The general target was a version that can be used in both single and double precision with as few changes  to the source code as possible. Our  goal was to achieve  a general setting of the working precision for all floating point variables at one location in one Fortran module. It has to be  taken into account that some parts of the code might require double precision. This fact was already noticed in the report on conversion of the IFS model by \citet{Vana2017}.
 
We will from now on refer to the single precision version as \emph{sp}, and to the double precision version as \emph{dp} version.

\subsection{Use of a precision switch}
\label{sec:switch}
One ideal and elegant way to switch easily between different precisions of the variables of a code in Fortran is to use a specification of the \texttt{kind} parameter for floating point variables as showed in the following example.
For reasons of flexibility, the objective of our work was  to have a radiation with such precision switch.
\begin{quote}
\begin{verbatim}
! define variable with prescribed working precision (wp):
real(kind = wp) :: x
\end{verbatim}
\end{quote}
The actual value of \texttt{wp} can then be easily switched in the following way:
\begin{quote}
\begin{verbatim}
! define different working precisions:
integer, parameter :: sp = 4        ! single precision (4 byte)
integer, parameter :: dp = 8        ! double precision (8 byte)
! set working precision:
integer, parameter :: wp = sp
\end{verbatim}
\end{quote}
The recommendation mentioned by \citet[Section 2.6.2]{metcalf2018} is to define the different values of the \texttt{kind} parameter by using the \texttt{selected\_real\_kind} function. It sets the actually used precision via the definition of the 
desired number of significant decimals (i.e., mantissa length) and exponent range, depending on the options the machine and compiler offer. This reads as follows:
\begin{quote}
\begin{verbatim}
! define precision using significant decimals and exponent range:
integer :: sign_decimals = 6
integer :: exp_range = 37
integer, parameter :: sp = selected_real_kind(sign_decimals, exp_range)   
...
integer, parameter :: dp = selected_real_kind(...,...)   
! set working precision:
integer, parameter :: wp = sp
\end{verbatim}
\end{quote}
In fact, similar settings can be found in the  ECHAM  module \texttt{rk\_mo\_kind},
but unfortunately they are not consequently used. Instead,    \texttt{kind = dp} is used directly in several modules.
A somehow dirty workaround, namely  assigning the value $4$ to  \texttt{dp} and declaring an additional precision for actual \emph{dp} where needed, circumvents this problem.
Then, compilation was possible  after some modifications (concerning MPI and NetCDF libraries and the module \texttt{mo\_echam\_radkernel\_cross\_messages}). The compiled code was crashing at runtime because of internal bugs triggered by code in the module \texttt{rk\_mo\_srtm\_solver} and other parts. These issues were solved later when investigating each code part with the incremental conversion method.
The cause of these bugs could not be easily tracked.
\subsection{Source code conversion of most time-consuming subroutines}
\begin{figure}[t]
\includegraphics[width=8.3cm]{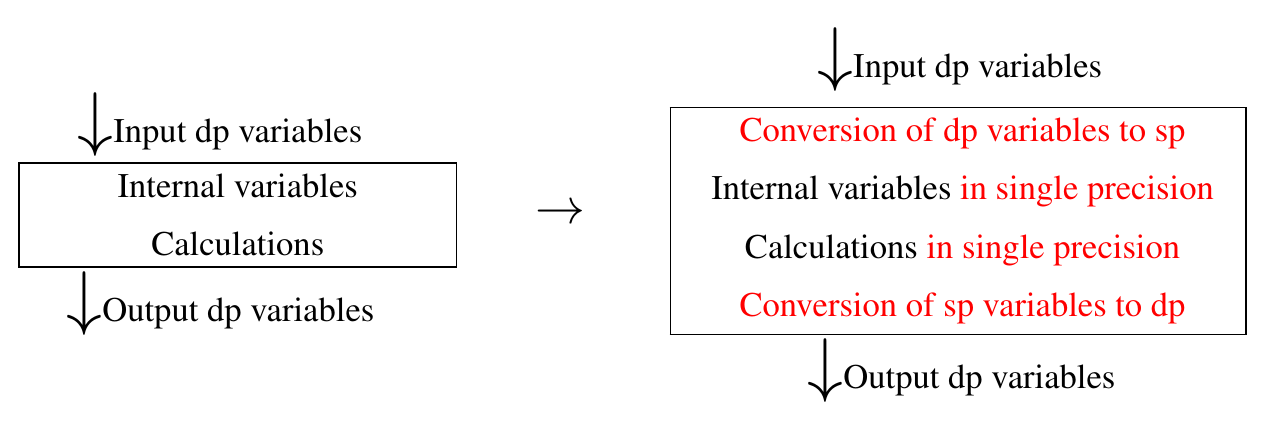}
 \caption{Necessary code changes to convert a subroutine/function from the original double precision version (left)  to single precision (right) with internal casting, modifications in red.}
 \label{fig:conversionscheme}
\end{figure}

As mentioned above, the conversion of the whole ECHAM model code using a simple switch  was not successful. Thus, we started to identify the most time-consuming subroutines and functions and converted them by hand.
This required the  conversion of input and output variables in the beginning and at the end of the respective subroutines and functions. The changes in the code are schematically depicted in Figure \ref{fig:conversionscheme}.

This procedure allowed an effective implementation of \emph{sp} computations of the converted subroutines/functions. We obtained  high performance gain in some code parts. But, the casting overhead destroyed the overall performance, especially if there are many variables to be converted.

\begin{figure}[t]
\includegraphics[width=8.3cm]{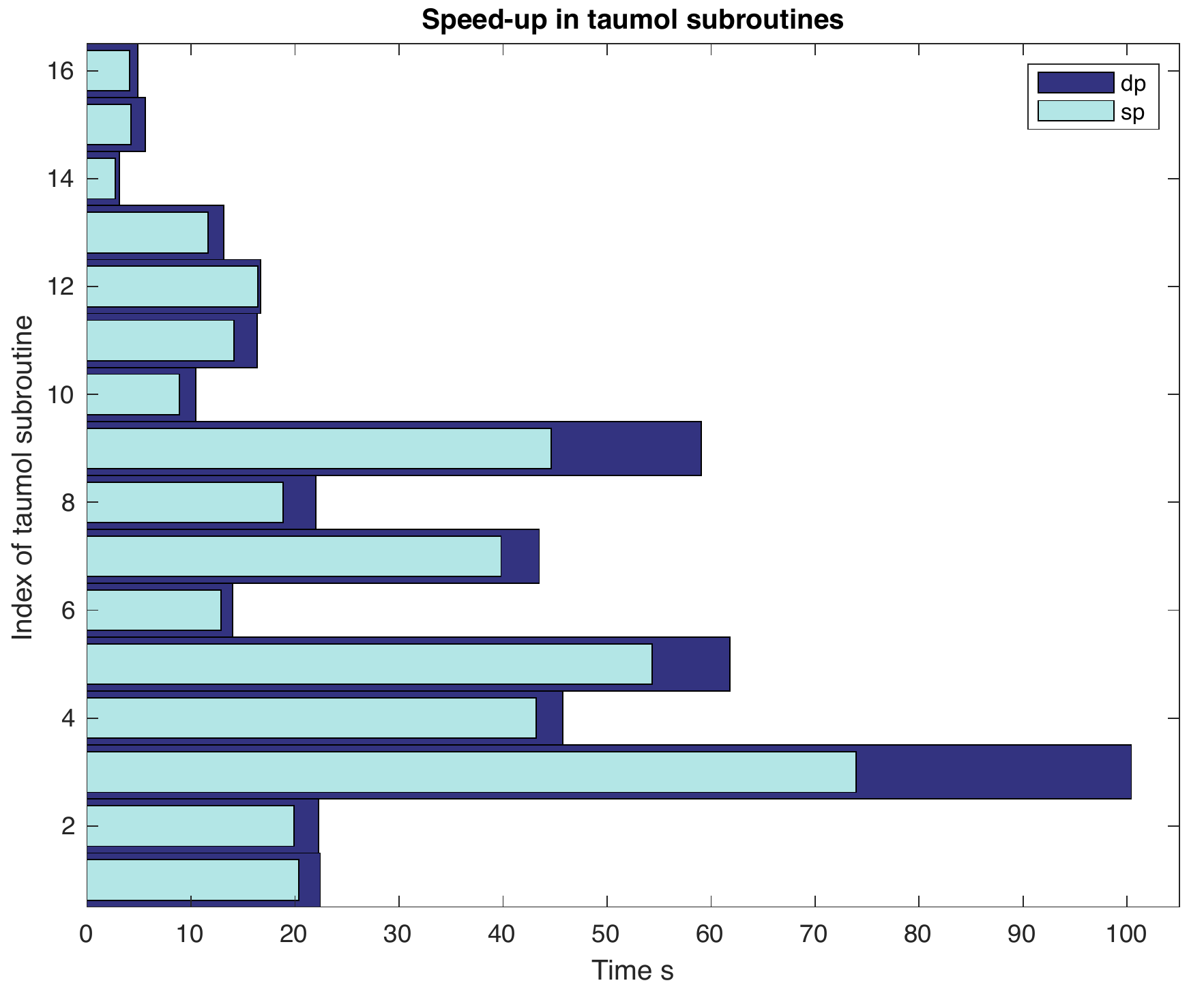}
\caption{Time consumption of single and double precision \texttt{taumol} subroutines.
   \label{fig:taumol}}
\end{figure}

For example, a time-consuming part of the subroutine \texttt{gas\_optics\_lw} in the module \texttt{mo\_lrtm\_gas\_optics} was converted in the above way. 
The converted part contains calls to subroutines \texttt{taumol01} to \texttt{taumol16}, which were converted to \emph{sp}.   Figure \ref{fig:taumol}  shows the  speed-up for these subroutines, which was up to 30\%. 
But   the needed casting in the calling subroutine \texttt{gas\_optics\_lw} doubled the overall runtime  in \emph{sp}, compared to \emph{dp}.

The results of this evaluation lead to the following conclusion, which is not very surprising: The bigger the converted code block is with respect to the number of input and output variables, the lower the overhead for the casting will be in comparison to the gain in the calculations that are actually performed in \emph{sp}. This was the reason for the decision  to convert the whole radiation part of ECHAM, as it contains a relatively small amount of input/output variables. 

\subsection{Incremental conversion of the radiation part}
\label{sec:incremental}
As a result of the  not efficient conversion of  the most time-consuming subroutines or functions only,  we performed a  gradual conversion of the whole radiation code. 
For this purpose, we started from the lowest level of its calling tree, treating each subroutine/function separately.
Consider an original subroutine on a lower level,
\begin{quote}
\begin{verbatim}
subroutine low(x_dp)
real(dp) :: x_dp
\end{verbatim}
\end{quote}
using \emph{dp} variables.
We renamed it as  \texttt{low\_dp}  and made a copy in \emph{sp}:
\begin{quote}
\begin{verbatim}
subroutine low_sp(x_sp)
real(sp) :: x_sp
\end{verbatim}
\end{quote}
We changed the \emph{dp} version  such that it just calls its \emph{sp} counterpart, using  implicit type conversions before and after the call:
\begin{quote}
\begin{verbatim}
subroutine low_dp(x_dp)
real(dp) :: x_dp
real(sp) :: x_sp
x_sp = x_dp
call low_sp(x_sp)
x_dp = x_sp
\end{verbatim}
\end{quote} 
Now we repeated the same procedure with each subroutine/function that calls the original \texttt{low}, e.g.,
\begin{quote}
\begin{verbatim}
subroutine high(...)
call low(x_dp)
\end{verbatim}
\end{quote}
We again renamed it as \texttt{high\_dp}, generated an \emph{sp} copy \texttt{high\_sp},
and defined an interface block \citep[]{metcalf2018}:
\begin{quote}
\begin{verbatim}
interface low
   module procedure low_sp
   module procedure low_dp
end interface
\end{verbatim}
\end{quote}
In both \texttt{high\_dp} and  \texttt{high\_sp}, we could now call the respective version of the lower level subroutine  passing either \emph{sp} or \emph{dp} parameters. 
The use of the  interface simplified this procedure significantly. 

If the model output was acceptable, the \emph{dp} version on the lower level, \texttt{low\_dp}, as well as the interface were redundant, and we deleted them. Finally, the \emph{sp} version could be renamed as \texttt{low}. 

This procedure was repeated up to the highest level of the calling tree. It required a lot of manual work, but it allowed the examination of each modified part of the code, as well as a validation of the output data of the whole  model.

In the ideal case, this would have led directly to a consistent \emph{sp} version. Replacing then the \texttt{sp} by \texttt{wp}  in the code for a working precision that could be set to \texttt{sp} or \texttt{dp} in some central module, we would have ended up with a model version that has a precision switch. The next section summarizes the few parts of the code that needed extra treatment.

\section{Necessary changes in the radiation code}
\label{sec:changes}
Changing the floating point precision in the radiation code required some modifications that are described in this section. Some of them are related to the use of external libraries, some others to an explicitly used precision-dependent implementation.

\subsection{Procedure}
In the incremental conversion, the precision variables \texttt{dp} and \texttt{wp} that are both used in the radiation code were replaced with \texttt{sp}.
Then in the final version, \texttt{sp} was replaced by \texttt{wp}. 
With this modification, \texttt{wp} became a switch for the radiation precision.
As the original radiation contained several variables lacking explicit declaration of their precision, the respective format specifiers were added throughout.

A Fortran compiler option (namely \texttt{-real-size 32}) can avoid the last step by assigning \emph{sp} to variables without format specifier.
However, since the original model used both \emph{sp} and \emph{dp},  this option could lead to an overhead due to conversion in the \emph{dp} part.
Although a compilation of the program with custom option was possible for the \emph{sp} Fortran modules, this procedure led to a more complicated compilation procedure and was therefore discarded.
The format specifier \texttt{D0}, denoting \emph{dp}, was also replaced by \texttt{\_wp}.


\subsection{Changes needed for the use of the NetCDF library}
In the NetCDF library, the names of  subroutines and functions have different suffixes depending on the used precision. They are used in the  modules
\begin{quote}
\begin{verbatim}
rk_mo_netcdf, rk_mo_cloud_optics, rk_mo_lrtm_netcdf,
rk_mo_o3clim, rk_mo_read_netcdf77, rk_mo_srtm_netcdf.
\end{verbatim}
\end{quote}
In \emph{sp}, they have to be replaced by their respective counterparts
 to read the NetCDF data with the correct precision.
The script shown in Appendix \ref{app:script} performs these changes automatically.
This solution was necessary because  an  implementation using an interface was causing crashes  for unknown reasons. It is possible that further investigation could lead to a working interface implementation for these subroutines/functions also at this point of the code.

 \subsection{Changes needed for parallelization with MPI}
Several interfaces of the module \texttt{mo\_mpi} were adapted to support \emph{sp}. In particular 
\texttt{p\_send\_real}, \linebreak
\texttt{p\_recv\_real} 
and \texttt{p\_bcast\_real} 
were overloaded with \emph{sp} subroutines for the needed array sizes. These modifications did not affect the calls to these interfaces. 
No conversions are made in this module, so no overhead is generated.

\subsection{Changes needed due to data transfer to the remaining ECHAM}
In ECHAM, data communication between the radiation part and the remaining atmosphere is implemented in the module
\begin{quote}
\begin{verbatim}
mo_echam_radkernel_cross_messages
\end{verbatim}
\end{quote}
 through subroutines using both MPI and the coupling library YAXT \citep{DKRZ2013}.  Since it was not possible to have a mixed precision data transfer for both libraries, our solution was to double the affected subroutines to copy and send both \emph{sp} and \emph{dp} data.
An additional variable conversion before or after their calls preserves the needed precision.
Also in this case, interface blocks were used to operate with the correct precision.
The changed subroutines have the following prefixes: \texttt{copy\_echam\_2\_kernel}, \texttt{copy\_kernel\_2\_echam}, \texttt{send\_atm2rad} and \texttt{send\_rad2atm}. These modifications only affect the ECHAM model when used in the parallel scheme. They have  a negligible overhead.

 \section{Parts still requiring higher precision}
 \label{sec:higherprecision}
In the \emph{sp} implementation of the radiation code, some parts are still requiring higher precision to run correctly. These parts  and the reasons  are presented in this section.

\subsection{Overflow avoidance}
When passing from \emph{dp} to \emph{sp} variables, the maximum representable number decreases from $\approx10^{308}$ to $\approx10^{38}$. In order to avoid overflow that could lead to crashes, it is 
necessary to adapt the code to new thresholds. A similar problem could potentially occur for numbers which are too small (smaller than $\approx10^{-45}$).

As stated in the  comments in the original code of \texttt{psrad\_interface}, the following exponential needs conversion if not used in \emph{dp}:
\begin{quote}
\begin{verbatim}
!this is ONLY o.k. as long as wp equals dp, else conversion needed
cisccp_cldemi3d(jl,jk,krow) = 1._wp - exp(-1._wp*cld_tau_lw_vr(jl,jkb,6)) 
\end{verbatim}
\end{quote}
 One plausible reason for this is that the exponential is too big for the range of \emph{sp}.
Even though this line was not executed in the used configuration, we converted the involved quantities to \emph{dp}.
Since the  variable on the left-hand side of the assignment was transferred within few steps to code parts outside the radiation, no other code inside the radiation had to be converted into \emph{dp}.

Also module \texttt{rk\_mo\_srtm\_solver} contained several parts sensitive to the precision.
First of all, the following lines containing the hard coded constant $500$ could cause overflow as well:
\begin{quote}
\begin{verbatim}
exp_minus_tau_over_mu0(:) = inv_expon(MIN(tau_over_mu, 500._wp), kproma) 
exp_ktau              (:) =     expon(MIN(k_tau,       500._wp), kproma)
\end{verbatim}
\end{quote}
Here, \texttt{expon} and \texttt{inv\_expon} calculate the exponential and inverse exponential of a vector (of length \texttt{kproma} in this case). The (inverse) exponential  of a number close to $500$ is too  big (small) to be represented in \emph{sp}.
In the used configurations,  this line was not executed either. Nevertheless,  we replaced this value  by a constant   depending on  the used precision, see the script in Appendix \ref{app:script}.


\subsection{Numerical stability}

Subroutine \texttt{srtm\_reftra\_ec} of module \texttt{rk\_mo\_srtm\_solver}, described in \citet{Meador1979}, showed to be very sensitive to the precision conversion. 
In this subroutine, already a conversion to \emph{sp} of just one of most  internal variables separately was causing crashes.
We introduced wrapper code for this subroutine to maintain the \emph{dp} version. 
The time necessary for this overhead was in the range of 3.5 to 6 \% for the complete radiation and between 0.6 and 3\% for the complete ECHAM model. 

In subroutine \texttt{Set\_JulianDay} of the module  \texttt{rk\_mo\_time\_base}, the use of \emph{sp}  for the variable \texttt{zd}, defined by
\begin{quote}
\begin{verbatim}
zd = FLOOR(365.25_dp*iy)+INT(30.6001_dp*(im+1))  &
     + REAL(ib,dp)+1720996.5_dp+REAL(kd,dp)+zsec 
\end{verbatim}
\end{quote}
caused crashes at the beginning of some simulation years.
In this case, the relative difference between the \emph{sp} and the \emph{dp} representation of the variable \texttt{zd} is close to machine precision (in \emph{sp} arithmetic), i.e., the relative difference attains its maximum value. This  indicates that  code parts that use this variable afterwards are very sensitive to small changes in input data.
The code block was  kept in \emph{dp} by reusing existing typecasts, without adding new ones.
Thus, this did not increase the runtime. Rewriting the code inside the subroutine might improve the stability and avoid the typecasts completely.

\subsection{Quadruple precision}
The module \texttt{rk\_mo\_time\_base} also  contains some parts in quadruple (\texttt{REAL(16)}) precision in the subroutine 
\linebreak 
\texttt{Set\_JulianCalendar}, e.g.:
\begin{verbatim}
zb = FLOOR(0.273790700698850764E-04_wp*za-51.1226445443780502865715_wp)
\end{verbatim}
Here \texttt{wp} was set to \texttt{REAL(16)} in the original code. This high precision was needed 
to prevent roundoff errors because of the number of digits in the used constants. 
We did not change the precision in this subroutine.  But since we used  \texttt{wp} as indicator for the actual working precision, we replaced 
\texttt{wp} by \texttt{ap} (advanced precision) to avoid conflicts with the working precision in this subroutine.
 We did not need to implement any precision conversion, since all input and output variables are converted from and to integer numbers inside the subroutine anyway.

\section{Results}
\label{sec:results}
In this section, we present the results obtained with the \emph{sp} version of the radiation part of ECHAM.
We show three types of results, namely a comparison of the model output, the obtained  gain in runtime and finally the gain in energy consumption. 

The results  presented below were obtained with the AMIP experiment \citep{WCRP2019} by using the coarse  (CR, T031L47) or low  (LR, T063L47) resolutions of ECHAM. The model was configured with the \texttt{cdi-pio} parallel input-output option \citep{Kleberg2017}.
We used the following compiler flags \citep{Intel2017}, which are the default ones for ECHAM:
\begin{itemize}
\item
\texttt{-O3}:  enables aggressive optimization,
\item \texttt{-fast-transcendentals},
 \texttt{-no-prec-sqrt},
\texttt{-no-prec-div}: enable faster but less precise transcendental functions, square roots, and divisions,
\item \texttt{-fp-model source}: rounds intermediate results to source-defined precision,
\item \texttt{-xHOST}: generates instructions for the highest instruction set available on the compilation host processor,
\item\texttt{-diag-disable 15018}: disables diagnostic messages,
\item\texttt{-assume realloc\_lhs}: uses different rules (instead of those of Fortran 2003) to interpret assignments.
\end{itemize}
\subsection{Validation of model output}
To estimate the output quality of the \emph{sp} version, we compared its results with 
\begin{itemize}
\item the results of the original, i.e., the \emph{dp} version of the model
\item and observational data available from several datasets.
\end{itemize}
We computed  the difference between the outputs of the \emph{sp} and \emph{dp} versions and  the differences of both versions to the observational data. 
We compared the values of
\begin{itemize}
\item
 temperature (at the surface and at 2m height), using the CRU TS4.03 dataset \citep{Cru2019},
 \item
 precipitation (sum of large scale and convective in ECHAM), using the GPCP  data provided by the NOAA/OAR/ESRL PSD, Boulder, Colorado, USA  \citep{Adler2003}.
 \item cloud radiative effect (CRE at the surface 
and at the top of the atmosphere, the latter split into longwave and shortwave parts), using the 
 CERES EBAF datasets release 4.0  \citep{Loeb2018}.
 \end{itemize}
 In all  results presented below, we use the monthly mean of these variables as basic data. This is motivated by the fact that we are interested in long time simulation runs and climate prediction rather than in short-term scenarios (as for  weather prediction). Monthly means are directly available as output from ECHAM.

 All computations  have been performed with the use of the 
\emph{Climate Data Operators (CDO)} \citep{Schulzweida2019}.

\subsubsection{Difference in RMSE between single and double version and observational data}
We computed the spatial root mean square error (RMSE) of the monthly means for both \emph{sp} and \emph{dp} versions and the above variables.  
We applied the same metric for the difference between the outputs of the \emph{sp} and \emph{dp} versions. 
We  computed these values over  time intervals where observational data were available in the used datasets.
For temperature and precipitation, these were the years 1981-2010, for CRE  the years 2000-2010 or 2001-2010. In all cases, we started the computation in the year 1970, having a reasonable time interval as spin-up. 

\begin{figure*}[t]
\includegraphics[width=\textwidth]{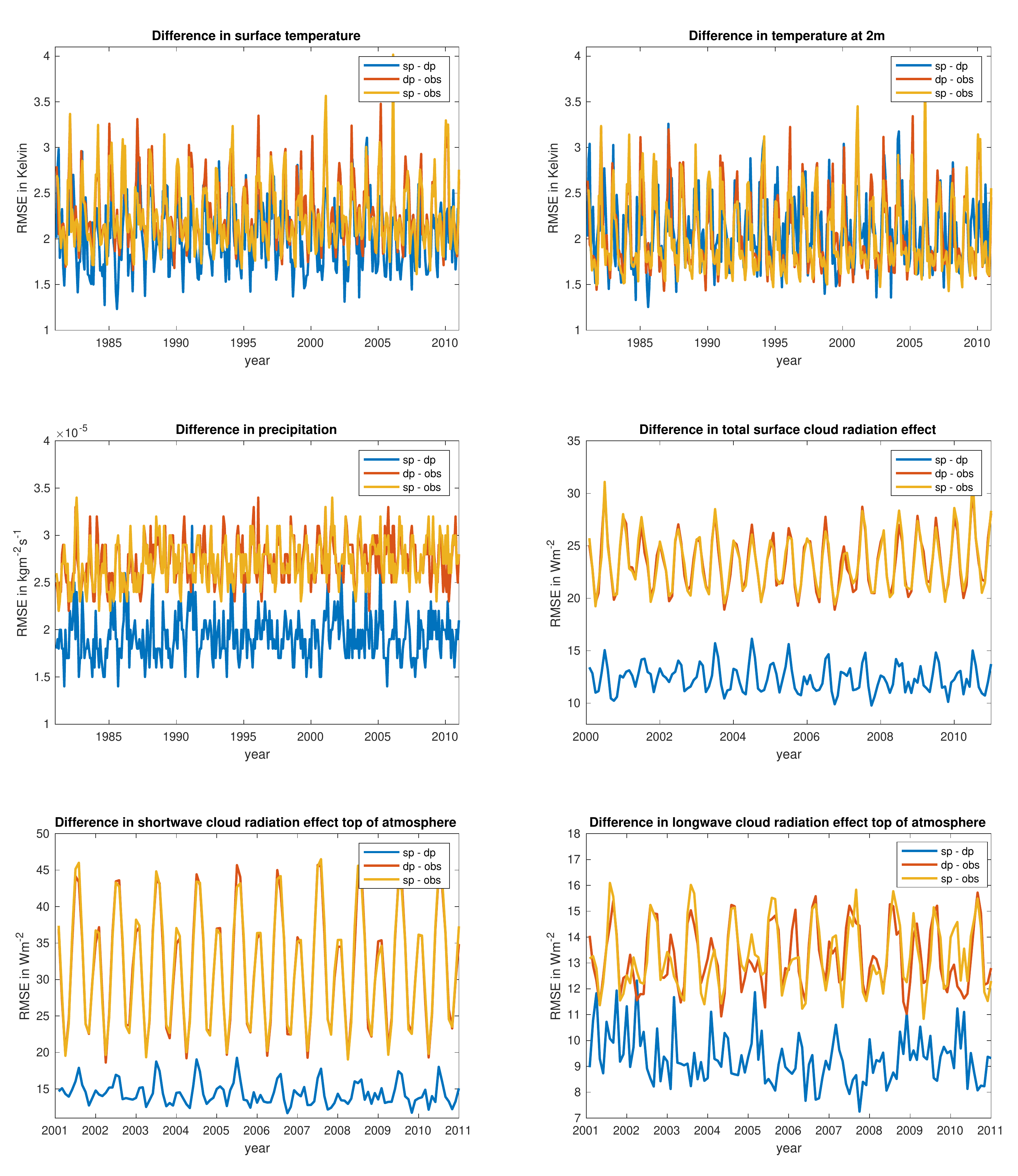}
\caption{Spatial RMSE of monthly  means for \emph{sp} and \emph{dp} versions and differences between them in the same metric, for (from top left to bottom right) temperature at the surface and at 2m, precipitation, total CRE at the surface, longwave and shortwave part of CRE at top of the atmosphere.
\label{fig:rmse-time}}
\end{figure*}

 Figure \ref{fig:rmse-time} shows the temporal behavior of the RMSE and the differences of \emph{sp} and \emph{dp} version, as they evolve in time. It can be seen that the RMSEs of  the \emph{sp} version  are  in the same magnitude as those  of the  \emph{dp} version. Also the differences between  both versions are of similar or even smaller magnitude.
Moreover, all RMSEs and  differences  do not grow in time. They oscillate but stay in the same order of magnitude for the whole considered time intervals.

Additionally, we averaged these values over the respective time intervals. Table \ref{table:rmse-mean} again shows  that the RMSEs of  the \emph{sp} version  are  in the same magnitude as those  of the  \emph{dp} version. Also the differences between  both versions are of similar or smaller magnitude.

\begin{table*}[t]
\begin{small}
\caption{Spatial RMSE of  monthly means 
for \emph{sp} and \emph{dp} versions and difference of both versions in the same metric, for selected model variables,  averaged over the respective time spans (obs = observational data).
\label{table:rmse-mean}}
\begin{tabular}{|c|c|c|c|c|c|c|}
\hline
Variable&ECHAM variable&Unit&Time span& \emph{sp} $-$ \emph{dp}& \emph{dp} $-$ obs &\emph{sp} $-$ obs\\
\hline
Surface temperature&169&K&1981 - 2010&2.0302&2.2862&2.2585\\
Temperature at 2m&167&K&1981 - 2010&2.0871&2.0585&2.0284\\
Precipitation&142+143&kgm${}^{-2}$s${}^{-1}$&1981 - 2010&$1.9281\cdot10^{-5}$&$2.7200\cdot10^{-5}$&$2.7161\cdot10^{-5}$\\
CRE, surface&176-185+177-186&Wm${}^{-2}$&2000 - 2010&12.3661&23.4345&23.4753\\
Shortwave CRE, &&&&&&\\
top of atmosphere&178-187&Wm${}^{-2}$&2001 - 2010&14.3859&31.5827&31.7220\\
Longwave CRE, &&&&&&\\
top of atmosphere&179-188&Wm${}^{-2}$&2001 - 2010&9.3010&13.2364&13.2903\\
\hline
\end{tabular}
\end{small}
\end{table*}

Since the RMSE  averages spatial differences, we also took a look at the minimum and maximum over all grid-points. This is a check  whether the conversion introduced arbitrary biases for the considered variables over the runtime. 
This method allows only a rough test, since a model producing unreliable data could anyway produce low averages, minima or maxima, even more if they are  averaged annually. 
 This comparison  showed a similar magnitude as the differences obtained using two subsequent ECHAM versions. Thus, we do not show corresponding plots here. 
 
Moreover, we compared our obtained differences  with the ones between two runs of the  ECHAM versions 6.3.02 and 6.3.02p1. The differences between \emph{sp} and \emph{dp} version are in the same magnitude as the differences between these two model versions.
\subsubsection{Spatial distribution of differences in the annual means}
We also studied the spatial distribution of the differences in the annual means. Again we considered the differences  between \emph{sp} and \emph{dp} version and of the output of both versions to the observations. Here we included the signs of the differences and no absolute values or squares.  For the given time spans, this results in a function of the form 
\begin{align*}
\text{DIFF}(\text{grid-point})
&:=
\frac{1}{\# \text{months in time-span}}
\sum_{\text{months in time-span}}
\big(y(\text{grid-point},\text{month})
-
z(\text{grid-point},\text{month})\big)
\end{align*}
for two variables or datasets $y,z$ of monthly data. This procedure can be used to see if some spatial points or areas are constantly warmer or colder over longer time ranges.
It is also a first test of the model output. However, it is clearly not sufficient for validation because errors may cancel out over time. 

\begin{figure*}[t]
\includegraphics[width=0.495\textwidth]{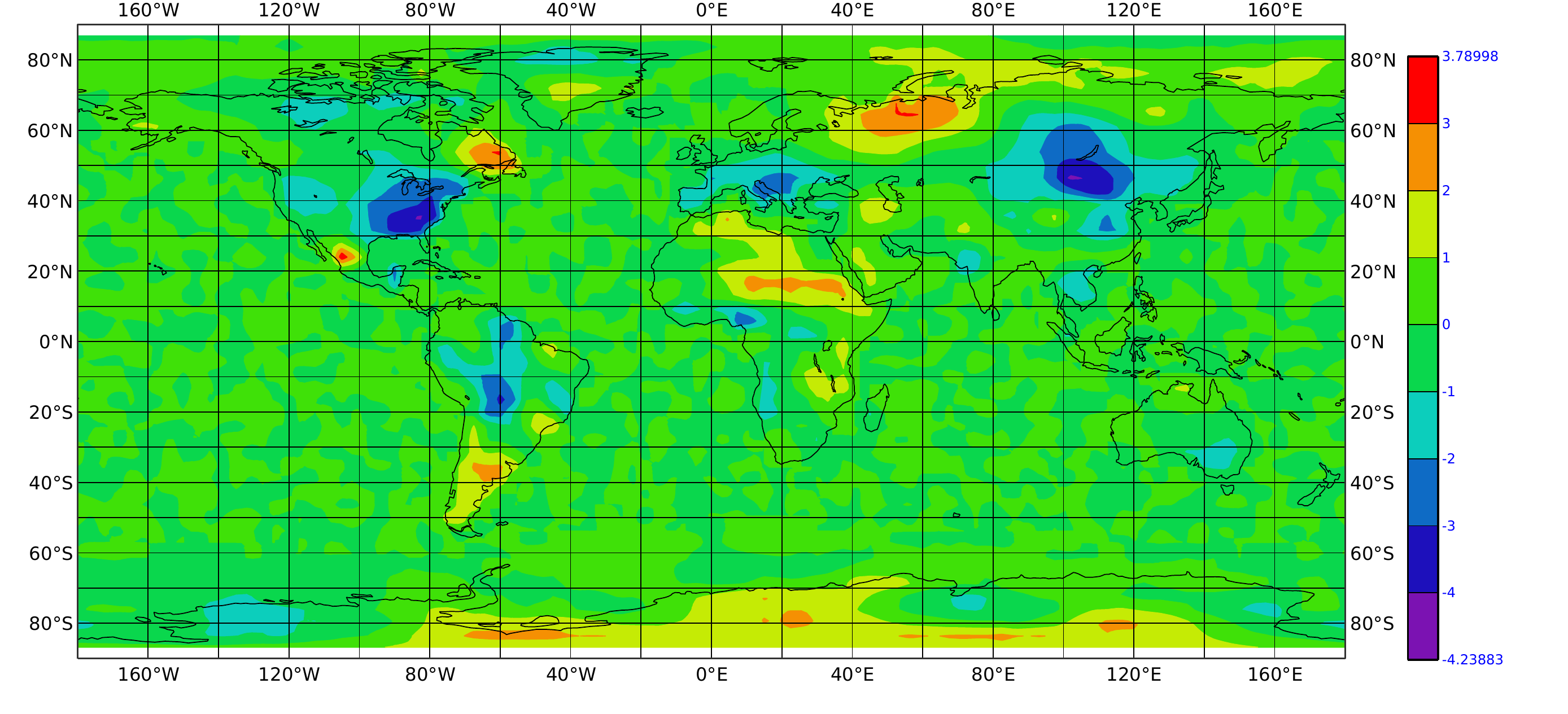}
\includegraphics[width=0.495\textwidth]{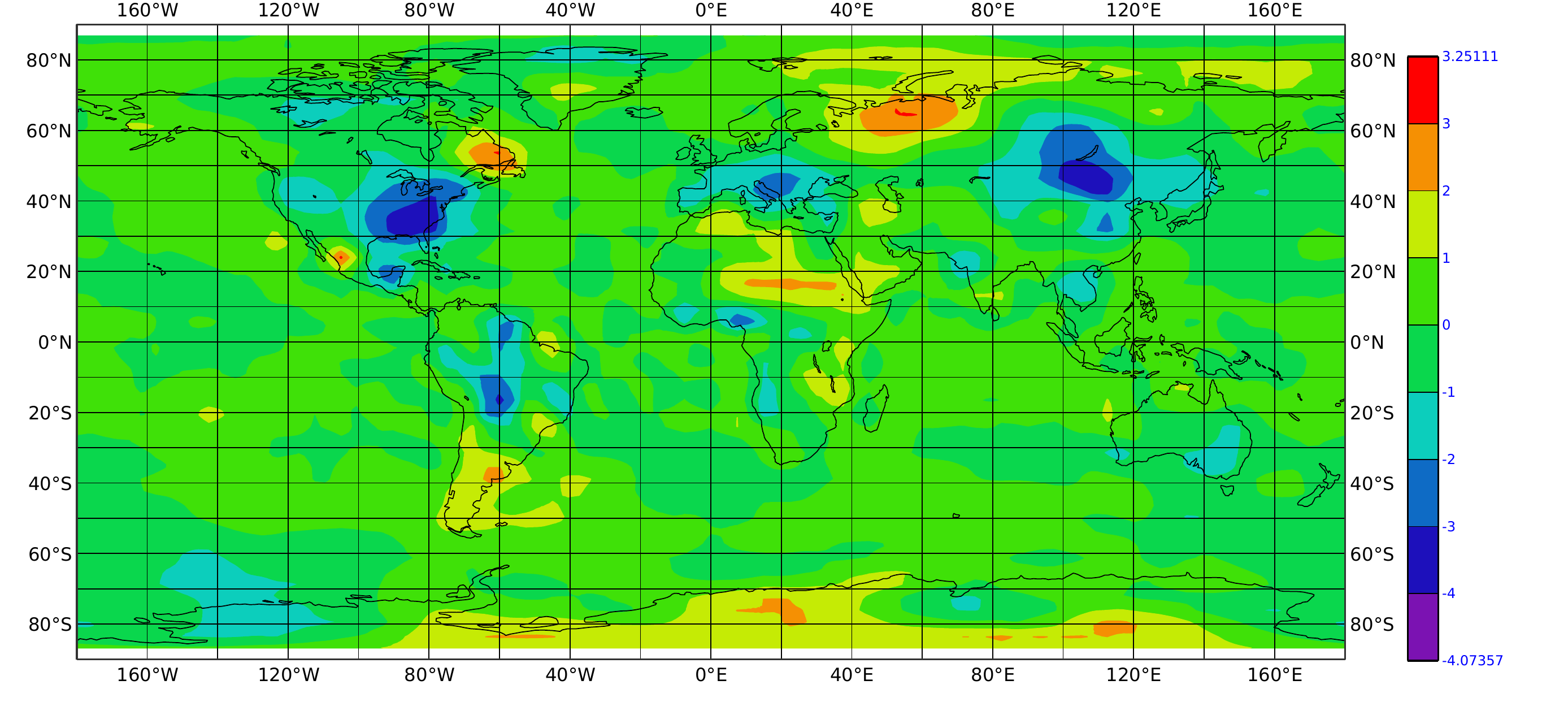}
\includegraphics[width=0.495\textwidth]{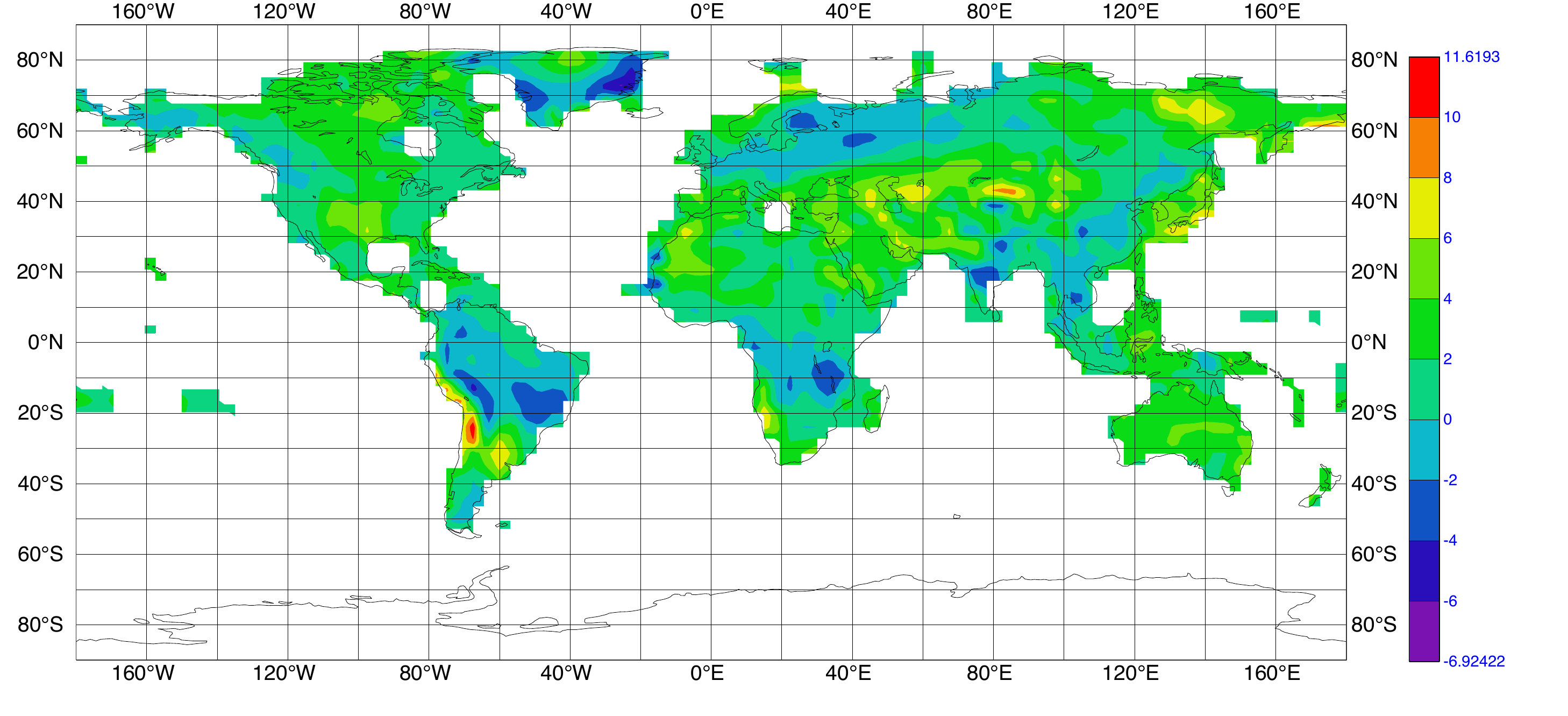}
\includegraphics[width=0.495\textwidth]{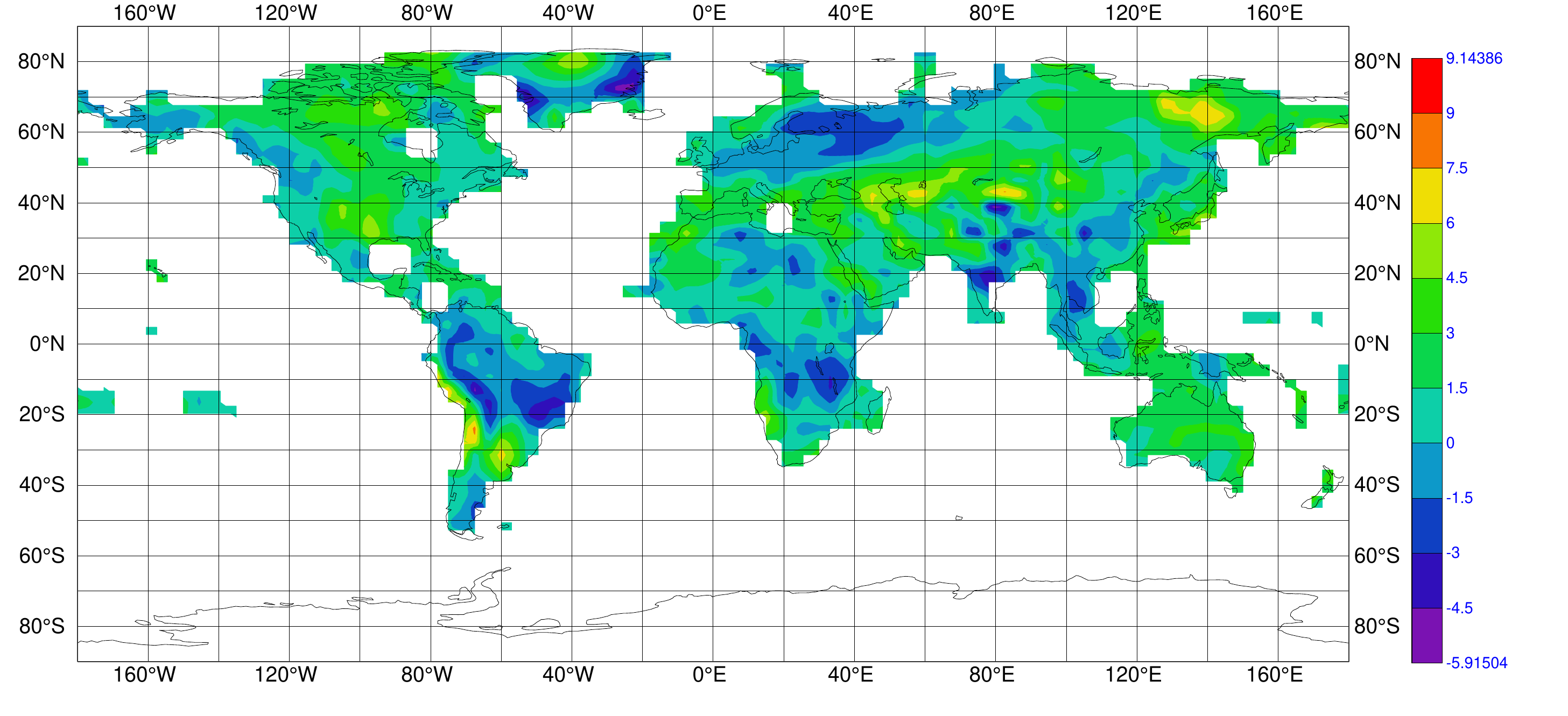}
\includegraphics[width=0.495\textwidth]{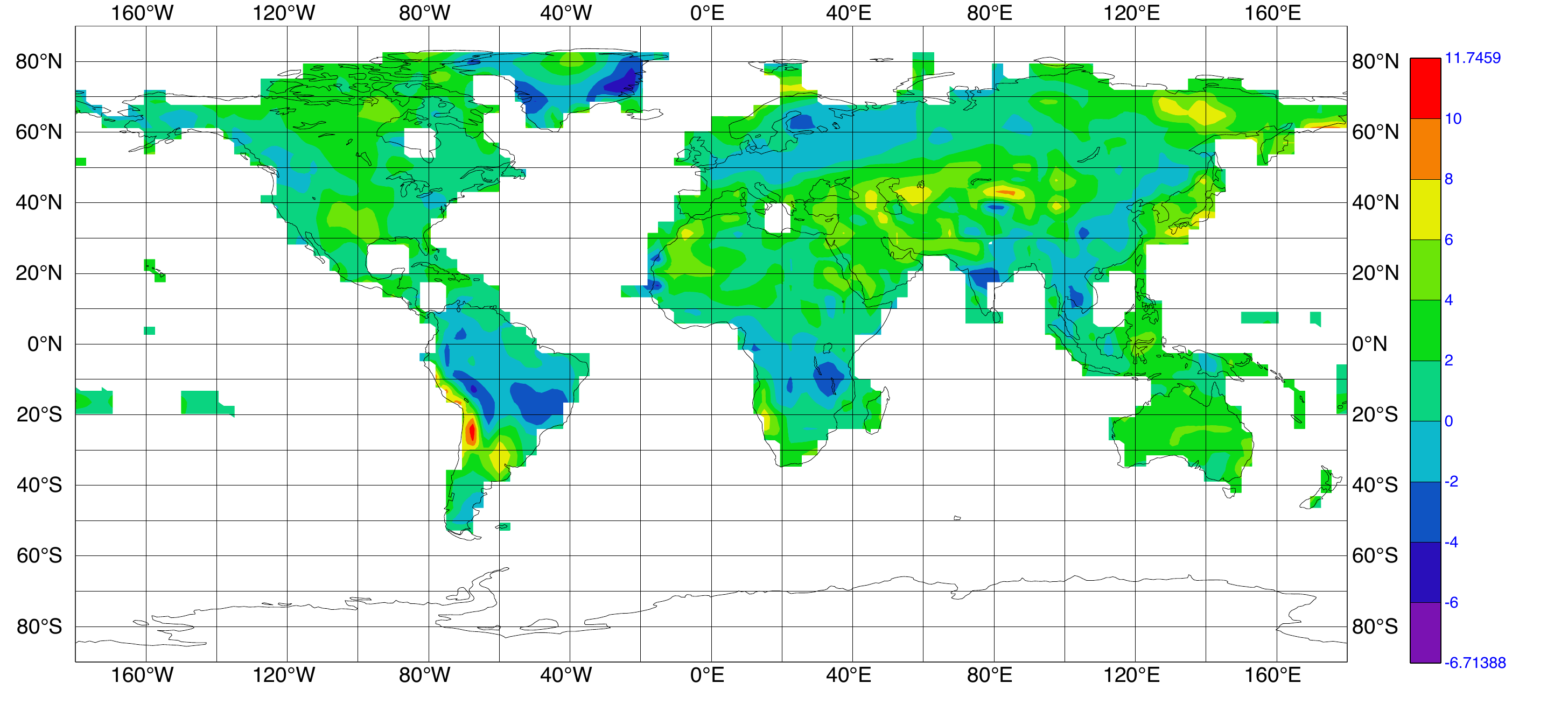}
\includegraphics[width=0.495\textwidth]{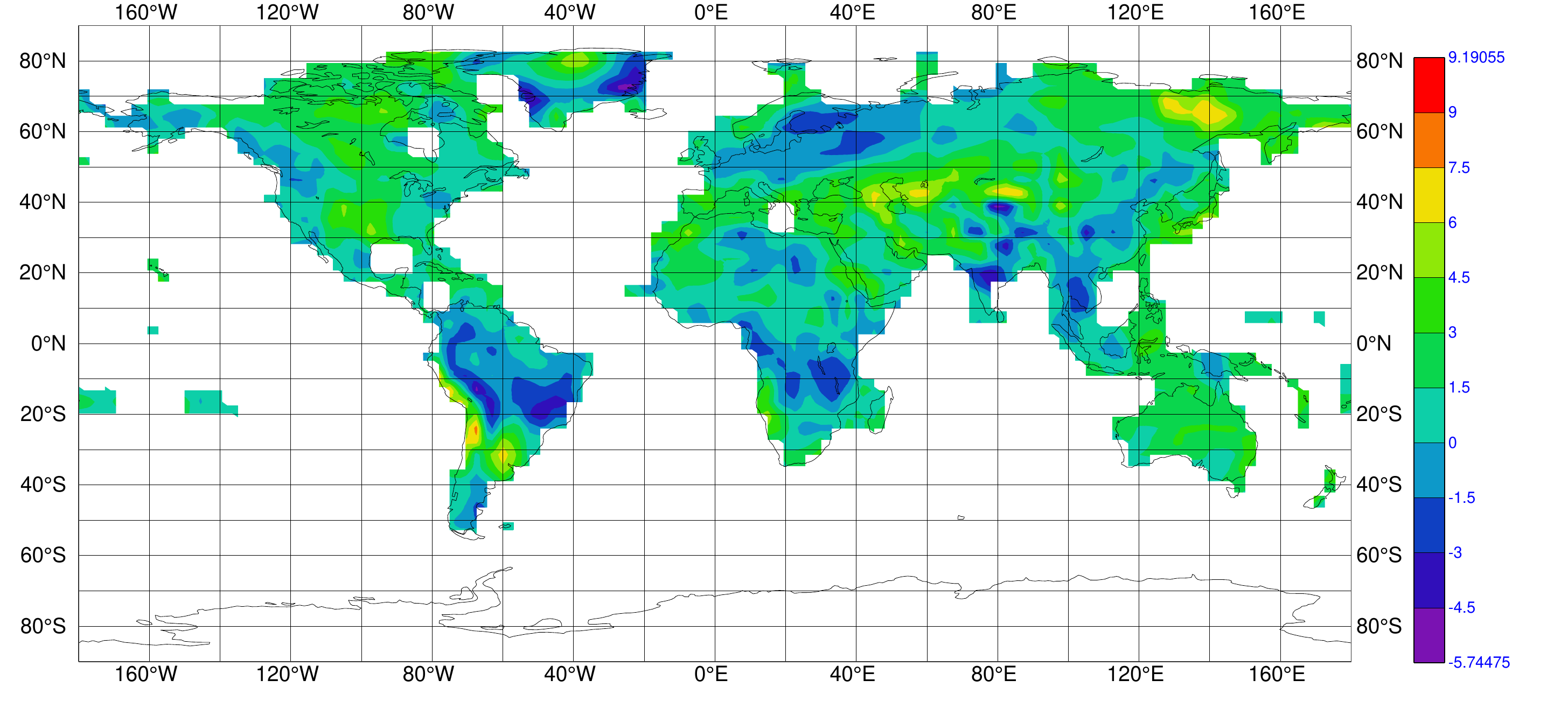}
\caption{Differences in temporal mean over  the time interval 1981 to 2010  in temperature at surface (left) and at 2m height (right) in Kelvin. Top:    two-sided t-test of hypothesis that the 30-years-mean of \emph{sp} version equals the one of the \emph{dp} version, computed with respect to variance of the annual \emph{sp} output, absolute values below 2.05 are not significant at the 95  confidence level. Middle row: difference between \emph{dp} version and observational data, bottom: between  \emph{sp} version and observational data. \label{fig:diff-time-temp}}
\end{figure*}

\begin{figure*}[t]
\includegraphics[width=0.495\textwidth]{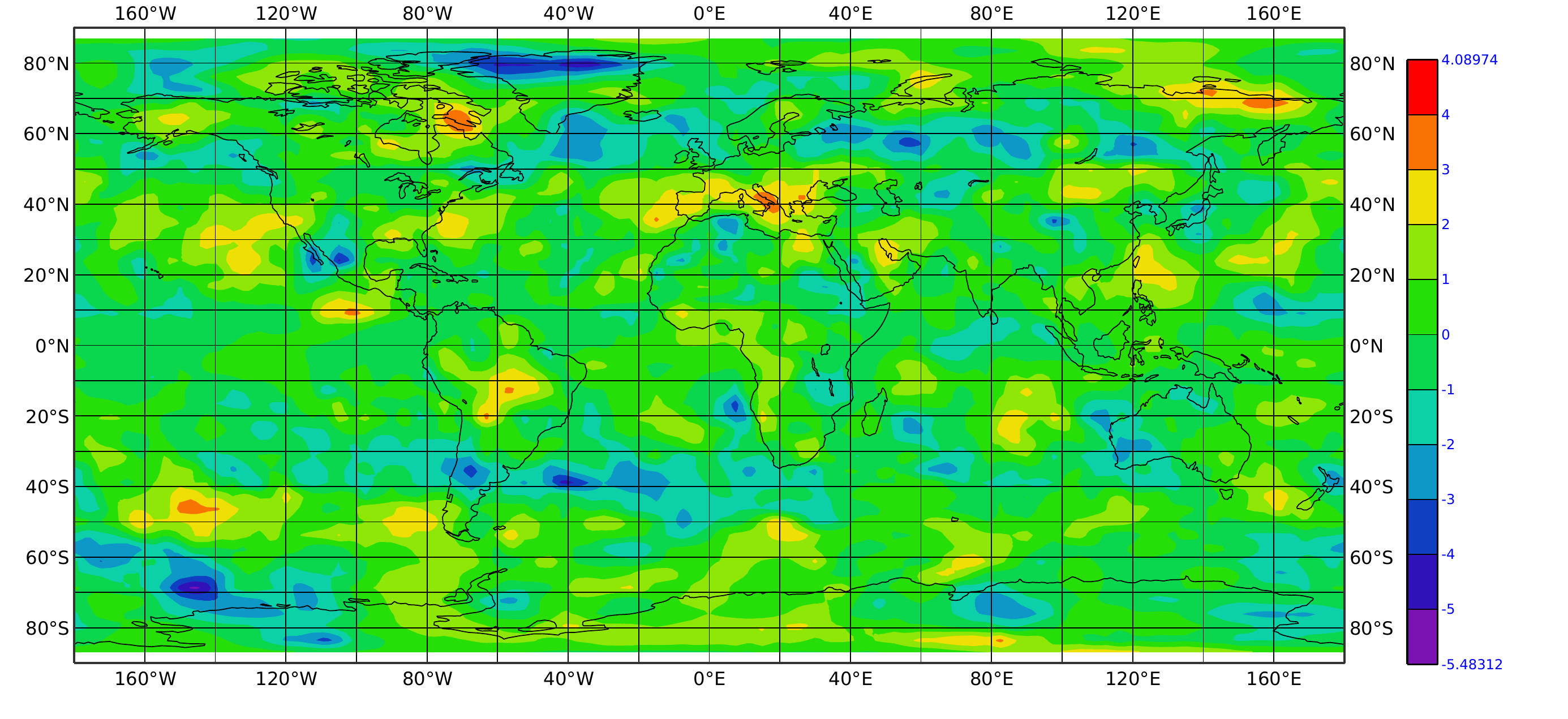}
\includegraphics[width=0.495\textwidth]{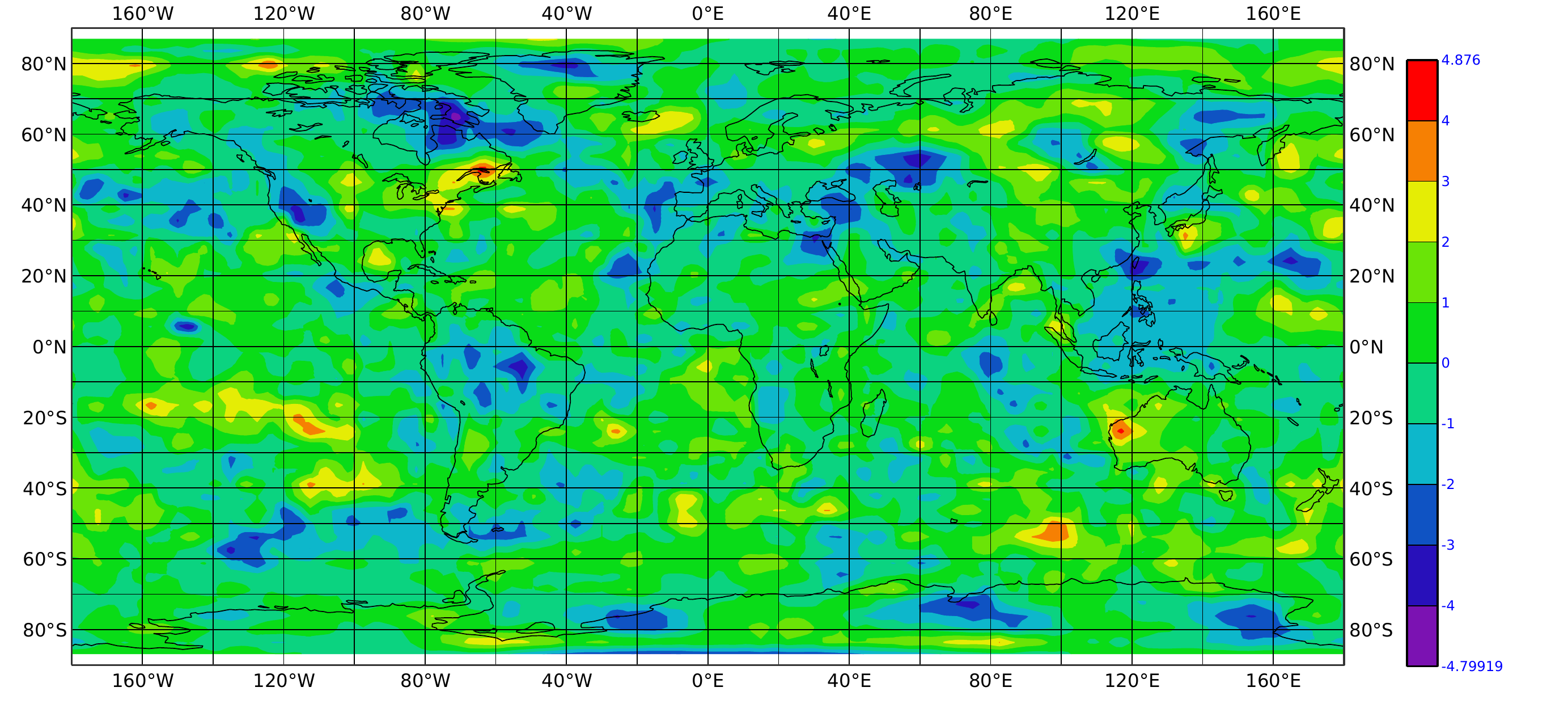}
\includegraphics[width=0.495\textwidth]{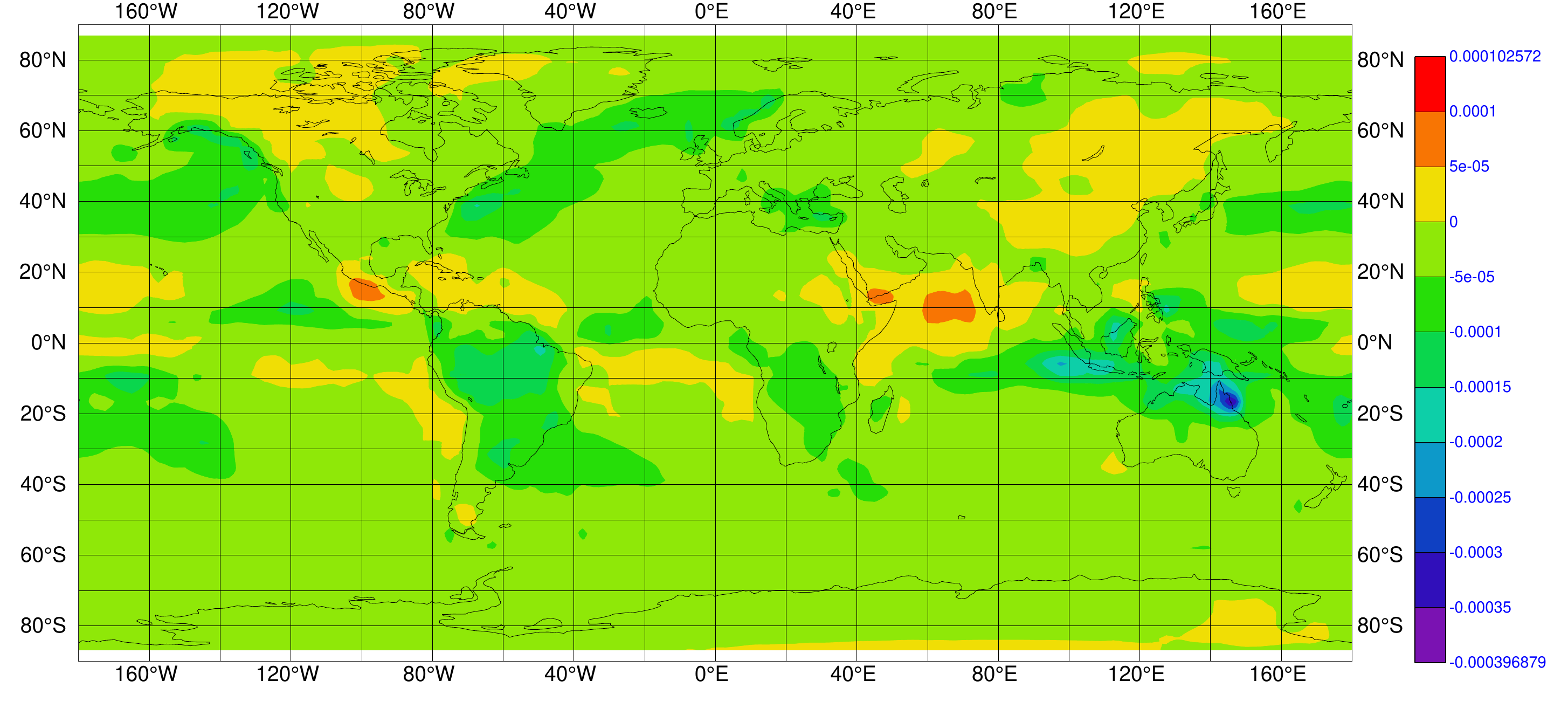}
\includegraphics[width=0.495\textwidth]{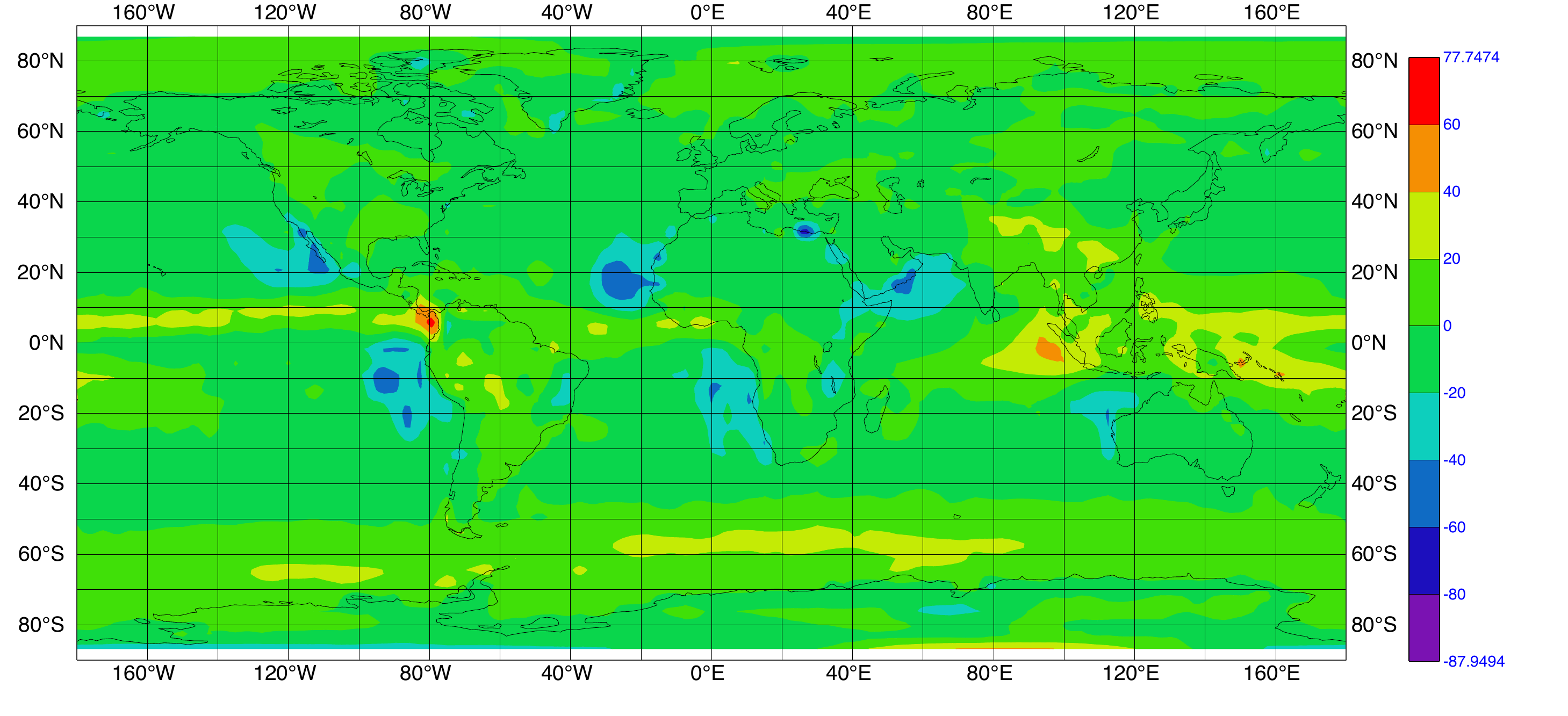}
\includegraphics[width=0.495\textwidth]{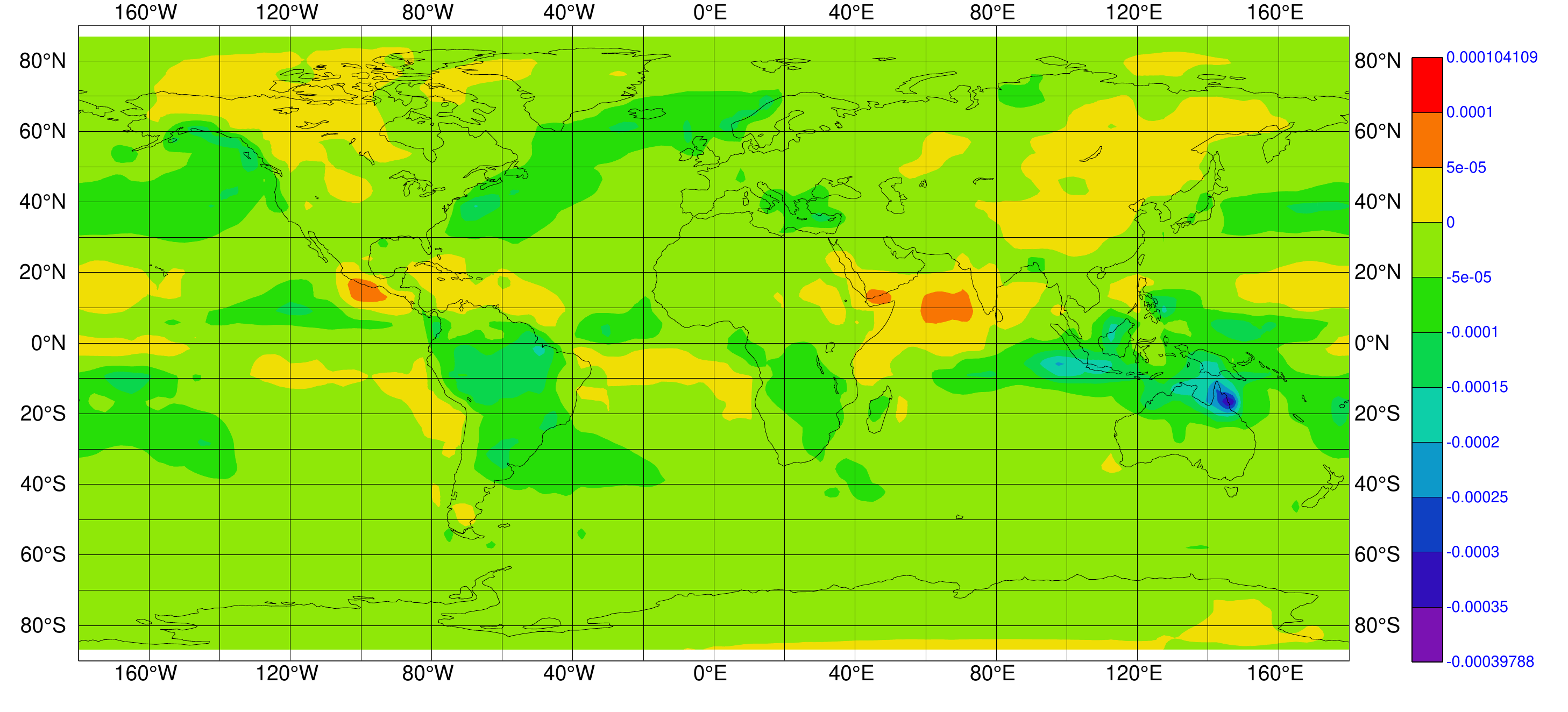}
\includegraphics[width=0.495\textwidth]{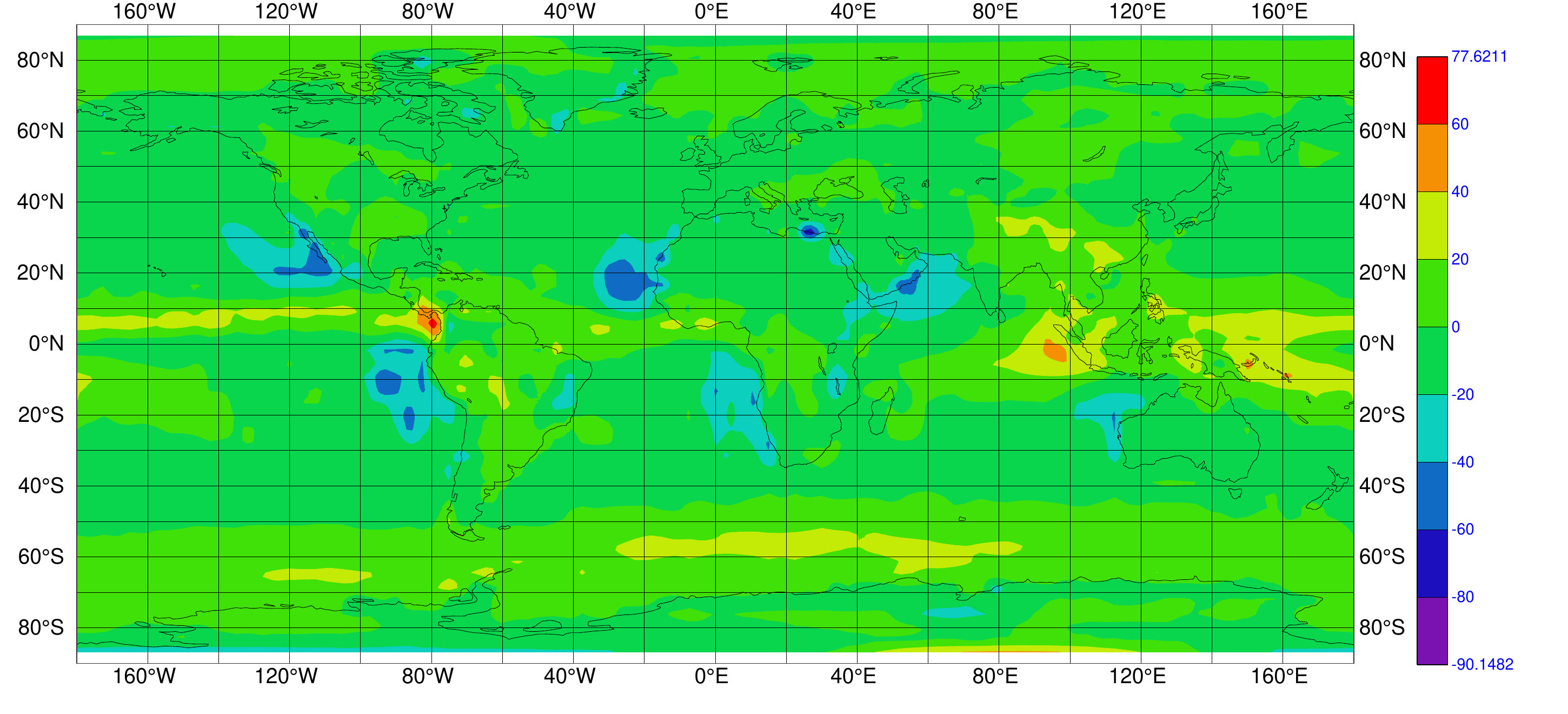}
\caption{As Figure \ref{fig:diff-time-temp}, but for  precipitation (left) in kgm${}^{-2}$s${}^{-1}$  and  surface cloud radiation effect  in Wm${}^{-2}$. For the latter difference to observations over 2000-2010. 
\label{fig:diff-time-precip-cre-surf}}
\end{figure*}

\begin{figure*}[t]
\includegraphics[width=0.495\textwidth]{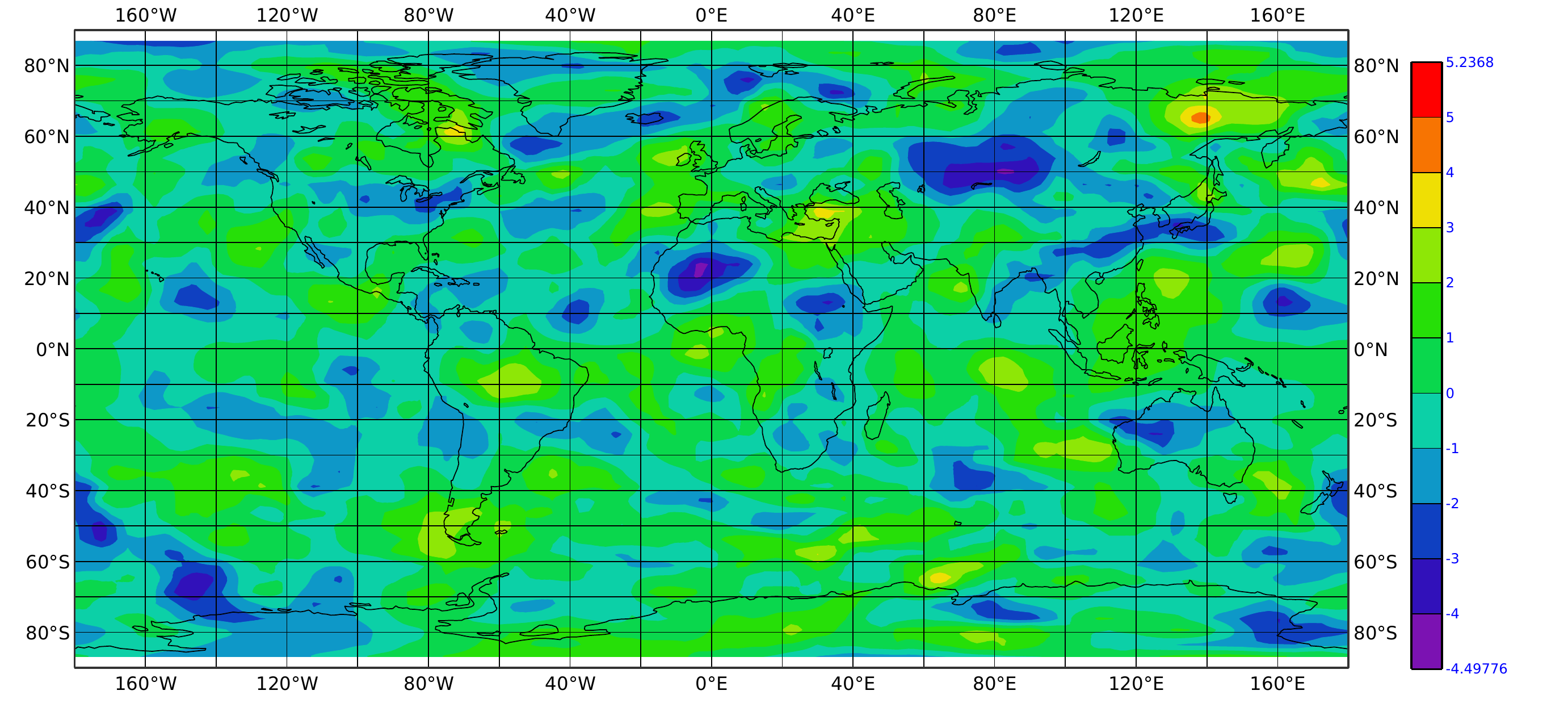}
\includegraphics[width=0.495\textwidth]{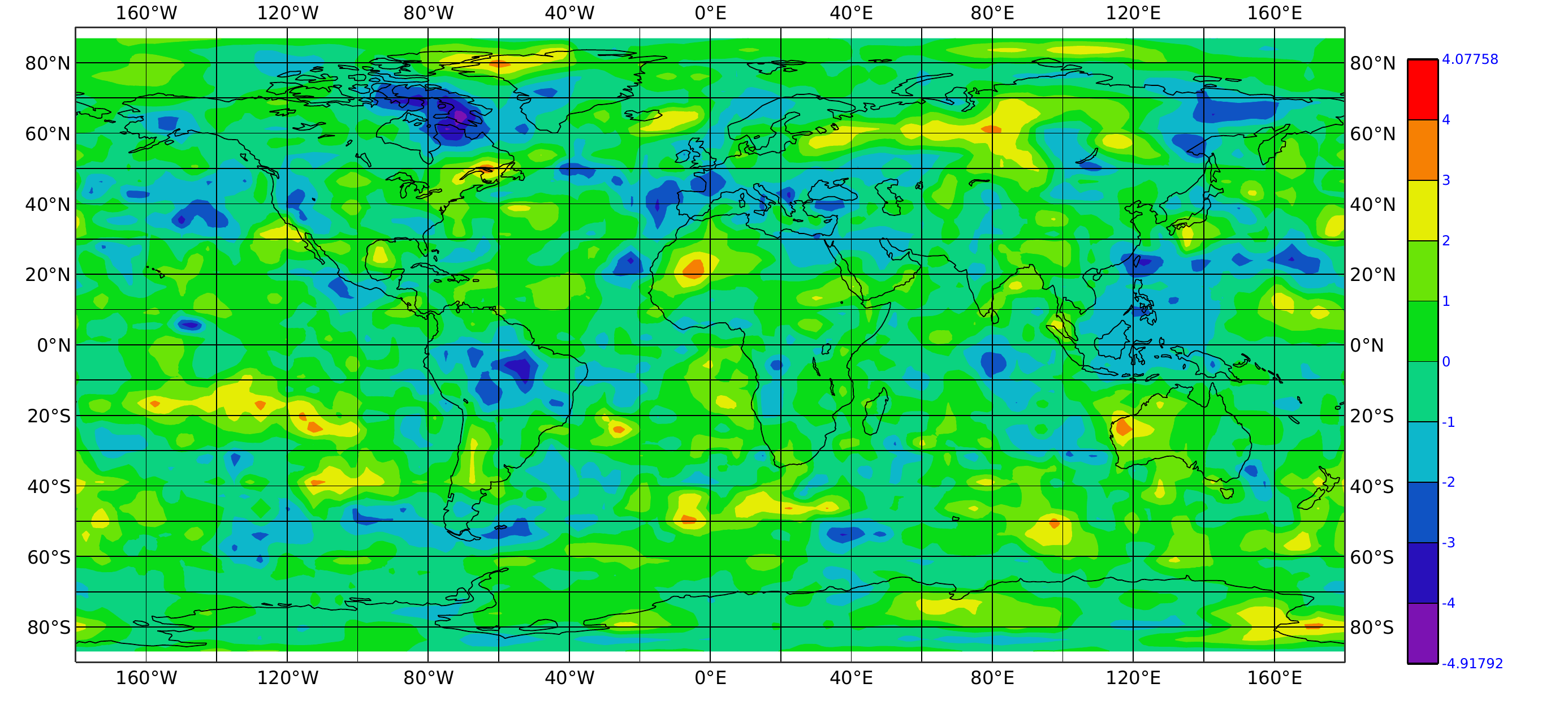}
\includegraphics[width=0.495\textwidth]{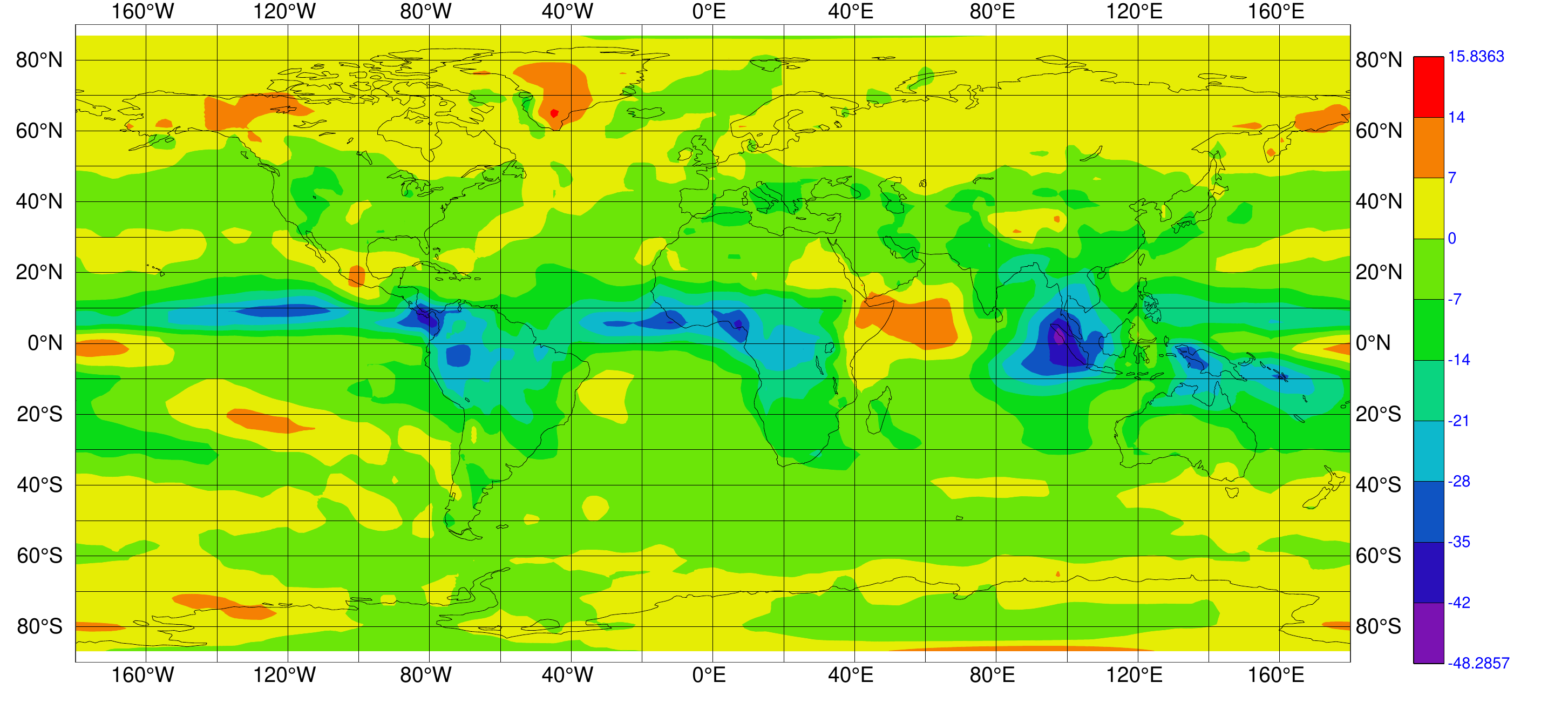}
\includegraphics[width=0.495\textwidth]{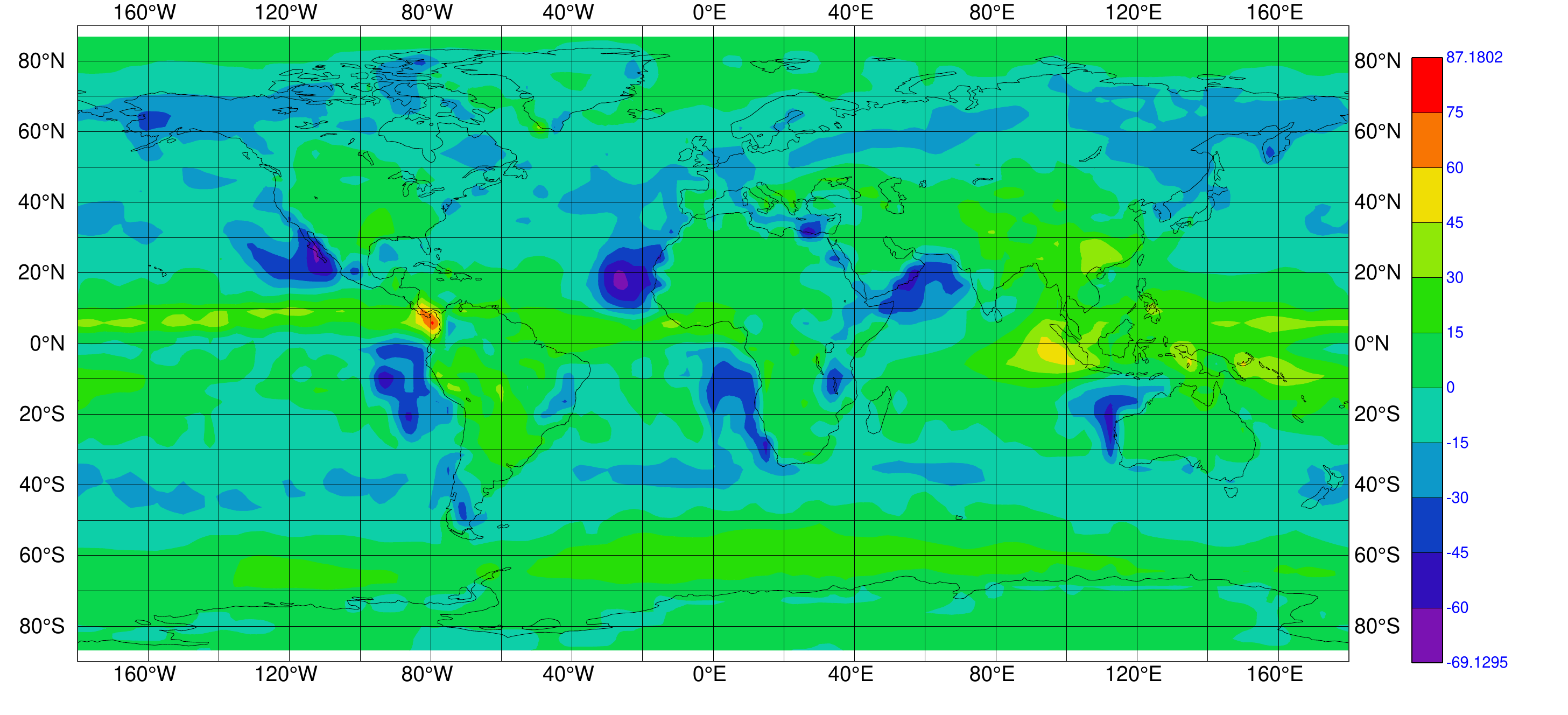}
\includegraphics[width=0.495\textwidth]{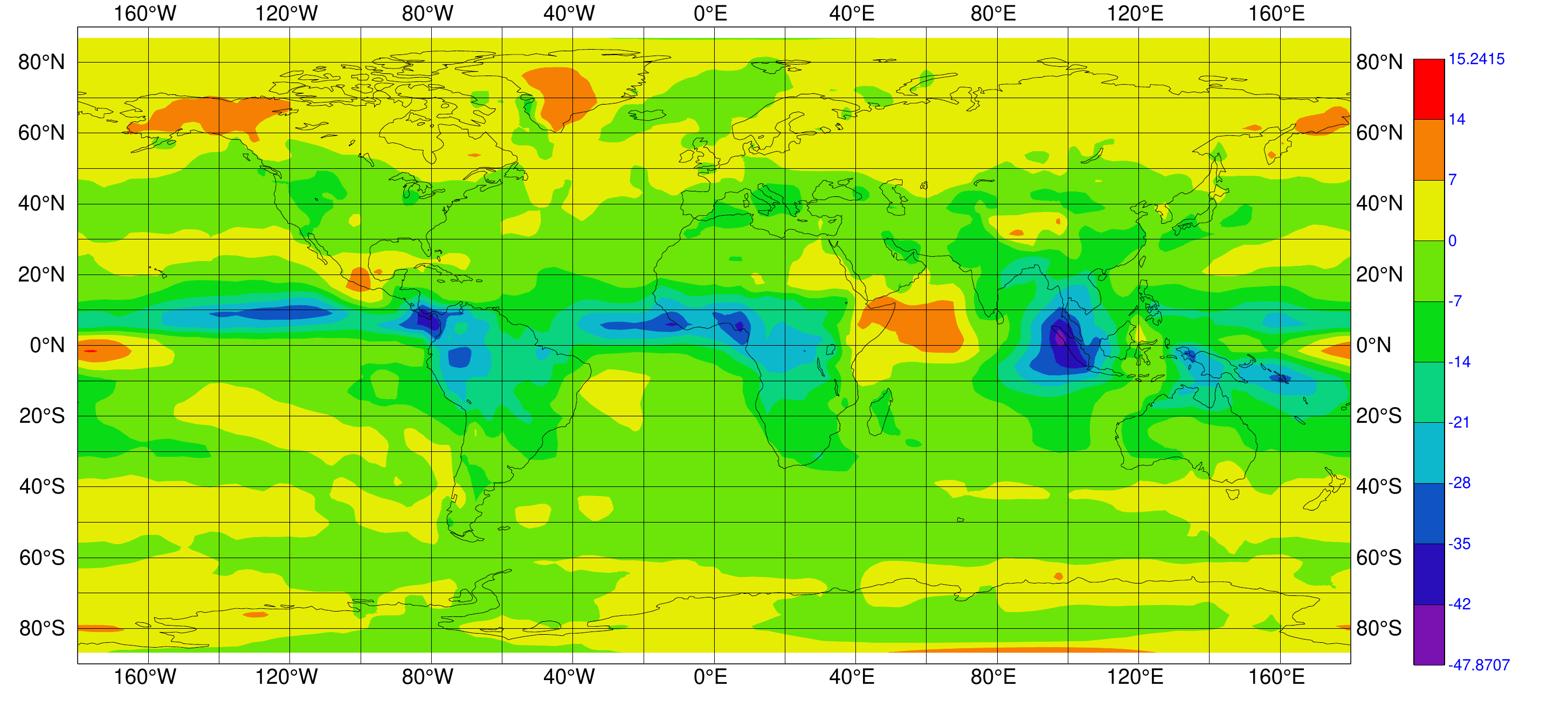}
\includegraphics[width=0.495\textwidth]{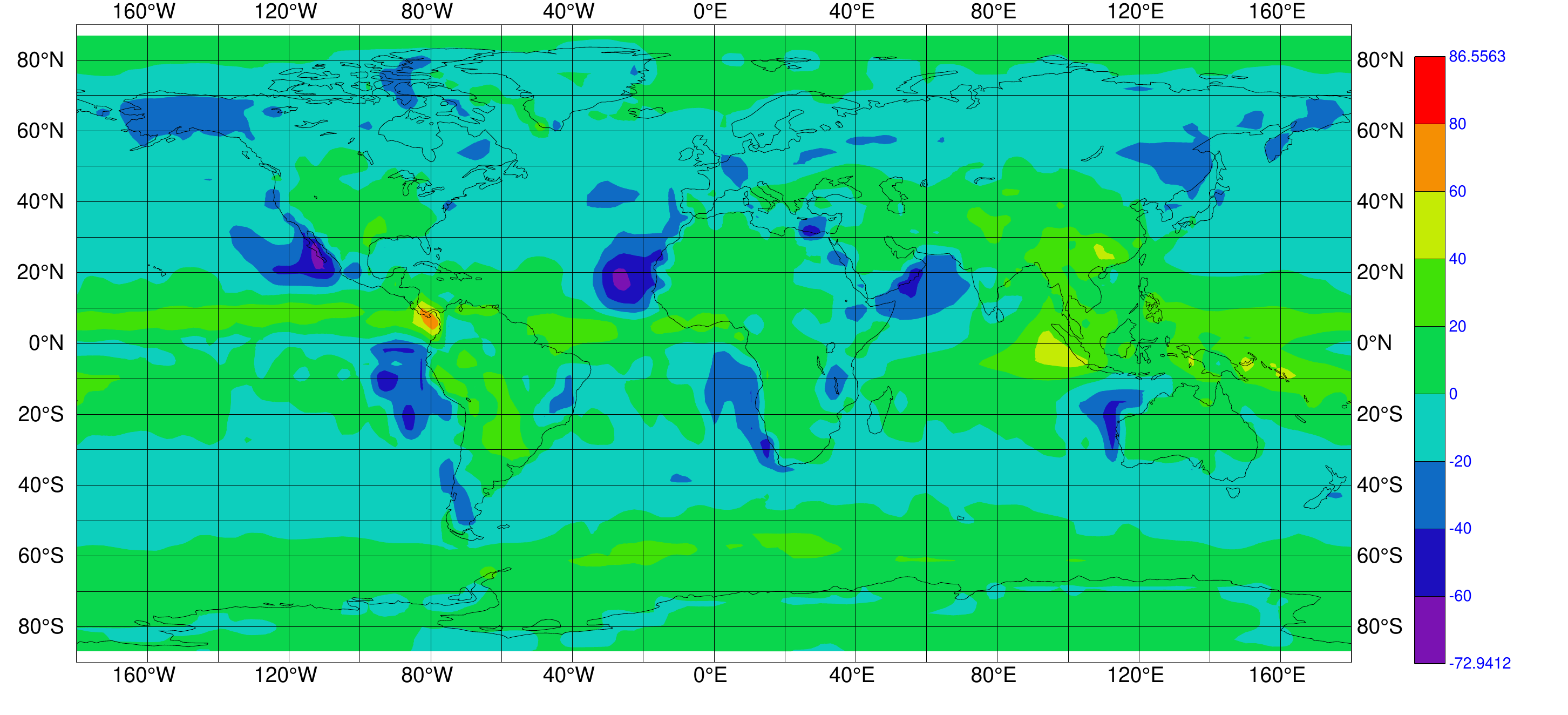}
\caption{As Figure \ref{fig:diff-time-temp}, but   longwave (left) and shortwave cloud radiation effect at the top of the atmosphere  in Wm${}^{-2}$. Difference to observations over  2001-2010 for both. \label{fig:diff-time-cre-toa}}
\end{figure*}

Here, we performed a statistic analysis of the annual means of the \emph{sp} version. We checked the hypothesis that the 30-years mean (in the interval 1981-2010) of the \emph{sp} version equals the one of the original \emph{dp} version. For this purpose, we applied a two-sided t-test, using a consistent estimator for the variance of the annual means of the \emph{sp} version. The corresponding values are showed in the respective top rows in  Figures \ref{fig:diff-time-temp} to \ref{fig:diff-time-cre-toa}. In this test, absolute values below 2.05 are not significant at the 95  confidence level. For all considered variables, it can be seen  that only very small spatial regions show higher values. 


\subsection{Speed-up}
In this section we present the results of the obtained speed-up when using the modified \emph{sp} radiation code in ECHAM. Since the model is usually run on parallel hardware, there are 
several configuration options that might affect its performance and also the speed-up when using the \emph{sp} instead of the \emph{dp} radiation code. We used the Mistral HPC system at DKRZ with 1 to 25 nodes, each of which has two Intel\textsuperscript{\textregistered} Xeon\textsuperscript{\textregistered} E5-2680v3 12C 2.5GHz (``Haswell") with 12 cores, i.e., using from 24 up to 600  cores.
The options we investigated are:
\begin{itemize}
\item The number of used nodes.
\item The choices \emph{cyclic:block} and \emph{cyclic:cyclic} (in this paper simply referred to as \emph{block} and \emph{cyclic})
 offered by the \emph{SLURM} batch system \citep{SchedMD2019}  used on Mistral. It controls the distribution of processes across nodes and sockets inside a node.
\item Different values of the ECHAM parameter \emph{nproma}, the maximum block length used for vectorization. For a detailed description see \citet[Section 3.8]{Rast2014}.
\end{itemize}

We were interested in the best possible speed-up when using the \emph{sp} radiation in ECHAM. 
We studied the performance gain achieved both for the radiation itself and for the whole ECHAM model for  a variety of different settings of the mentioned options, for both CR and (with reduced variety) LR 
 resolutions. Our focus lies on the CR version, since it is the configuration that is used in the long-time paleo runs intended in the PalMod project.
 
 The results presented in this section have been generated with  the \emph{Scalable Performance Measurement Infrastructure for Parallel Codes} \citep{ScoreP2019}
 and the internal ECHAM timer. 
 
All time measurements are based on one-year runs. The unit we use to present the results is the number of simulated years per day runtime. It  can be computed by the time measurements of the one-year runs. For the results for the radiation part only, these are  theoretical numbers, since the radiation is not run stand-alone for one year in ECHAM. We include them to give an impression what might be possible when more parts of ECHAM or even the whole model would be converted to \emph{sp}. Moreover, we wanted to see if the speed-up of 40\% achieved with IFS model in \citet{Vana2017} could be reached.

 To figure out if there are significant deviations in the runtime, we also applied a statistic analysis for  100 one-year runs. They showed that there are only very small relative  deviations from the mean, see Table \ref{tab:mean-runtime}.
 
\begin{table*}[t]
\caption{Relative standard deviations of runtime over 100 runs.
\label{tab:mean-runtime}}
\begin{tabular}{|r|c|c|c|c|}
\hline
 &24 nodes block  & 24 nodes cyclic &1 node block  & 1 node cyclic   \\
\hline
Radiation dp&0.0095&0.0104&0.0081&0.0075\\
sp&0.0132&0.0122&0.0079&0.0084\\
\hline
ECHAM dp&0.0220&0.0179&0.0030&0.0023\\
sp&0.0158&0.0189&0.0027&0.0020\\
\hline
\end{tabular}
\end{table*} 
 
At the end of this section, we give some details which parts of the radiation code benefit the most from the conversion to reduced precision, and which ones not.

\subsubsection{Dependency of runtime and speed-up  on parameter settings}

\begin{figure*}[t]
\includegraphics[width=\textwidth]{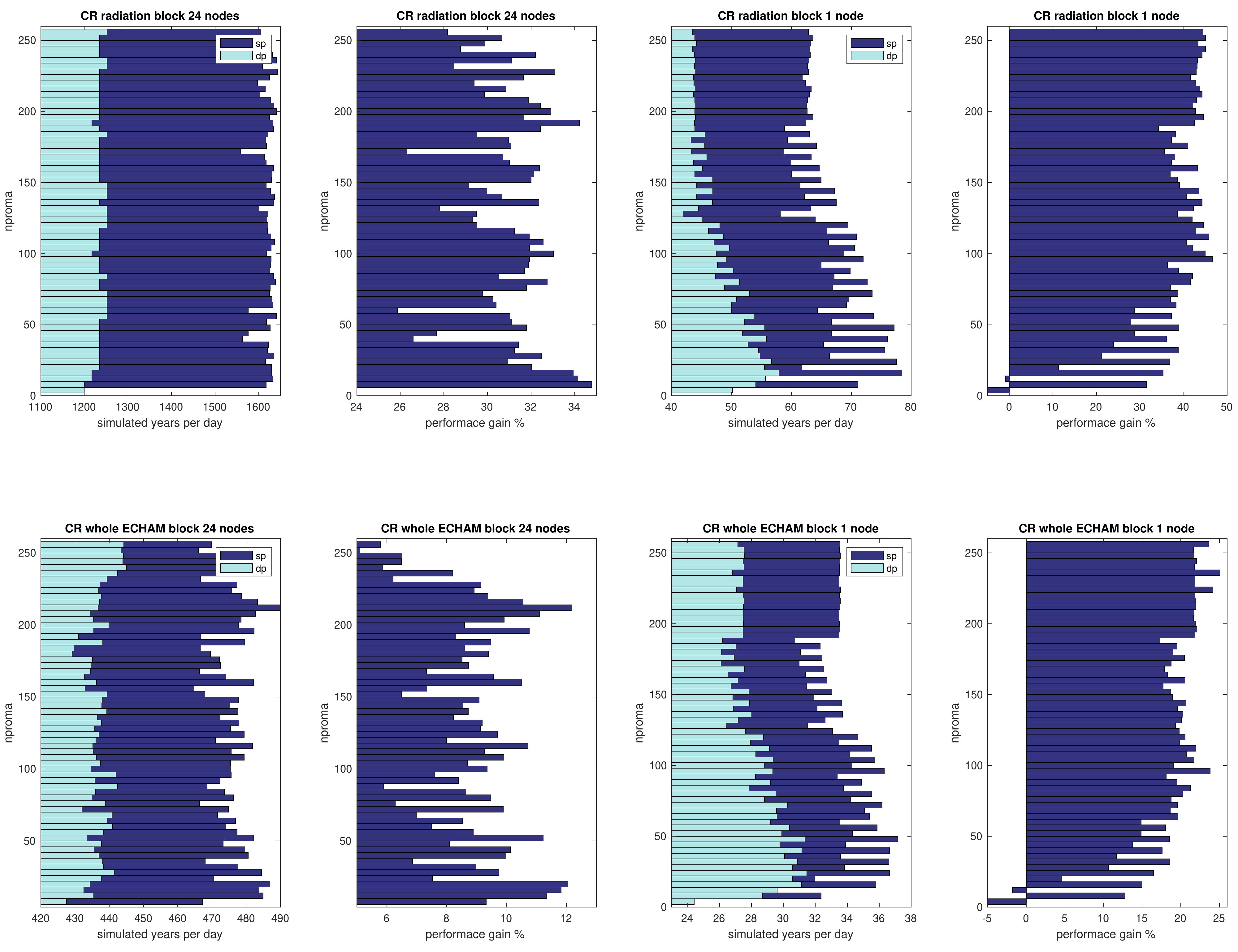}
\caption{Comparison of simulated model years per day cputime for \emph{sp} and \emph{dp} versions in coarse resolution (CR), for radiation part (top) and whole ECHAM (bottom), 1 and 24 nodes and  values of \emph{nproma} between 4 and 256, in steps of 4.
\label{fig:CR-nproma-block}}
\end{figure*}

In order to find out the best possible speed-up when using the \emph{sp} radiation code, we first analyzed the dependency of the runtime on the parameter \emph{nproma}. For the CR resolution, we tested for 1 to 25 cores  \emph{nproma} values from 4 to 256 in steps of 4.  It can be seen  in the two top left pictures in Figure \ref{fig:CR-nproma-block} that for 24 nodes there is no big dependency on \emph{nproma} for the original \emph{dp} version, when looking at radiation only. For the  \emph{sp} version, the dependency is slightly bigger, which results in a variety of the achieved speed-up between 25 to 35\%. 

When looking at the results for the whole ECHAM  model on the two left pictures below in in Figure \ref{fig:CR-nproma-block}, it can be seen that the dependency of the speed-up on \emph{nproma} becomes more significant.

 Using only one node the performance for the \emph{dp} version decreases with higher \emph{nproma}, whereas the \emph{sp} version does not show that big dependency. The effect is stronger when looking at the radiation time only than for the whole ECHAM. For very small values of \emph{nproma}, the \emph{sp} version was even slower than the \emph{dp} version. 
 In particular, the default parameter value (\emph{nproma} =  12) for the \emph{sp} version  resulted in slower execution time than the corresponding \emph{dp} version.
An increased value of the parameter (\emph{nproma} = 48) made \emph{sp} faster, even compared to the fastest \emph{nproma} for \emph{dp} (which was 24).

\begin{figure*}[t]
\includegraphics[width=\textwidth]{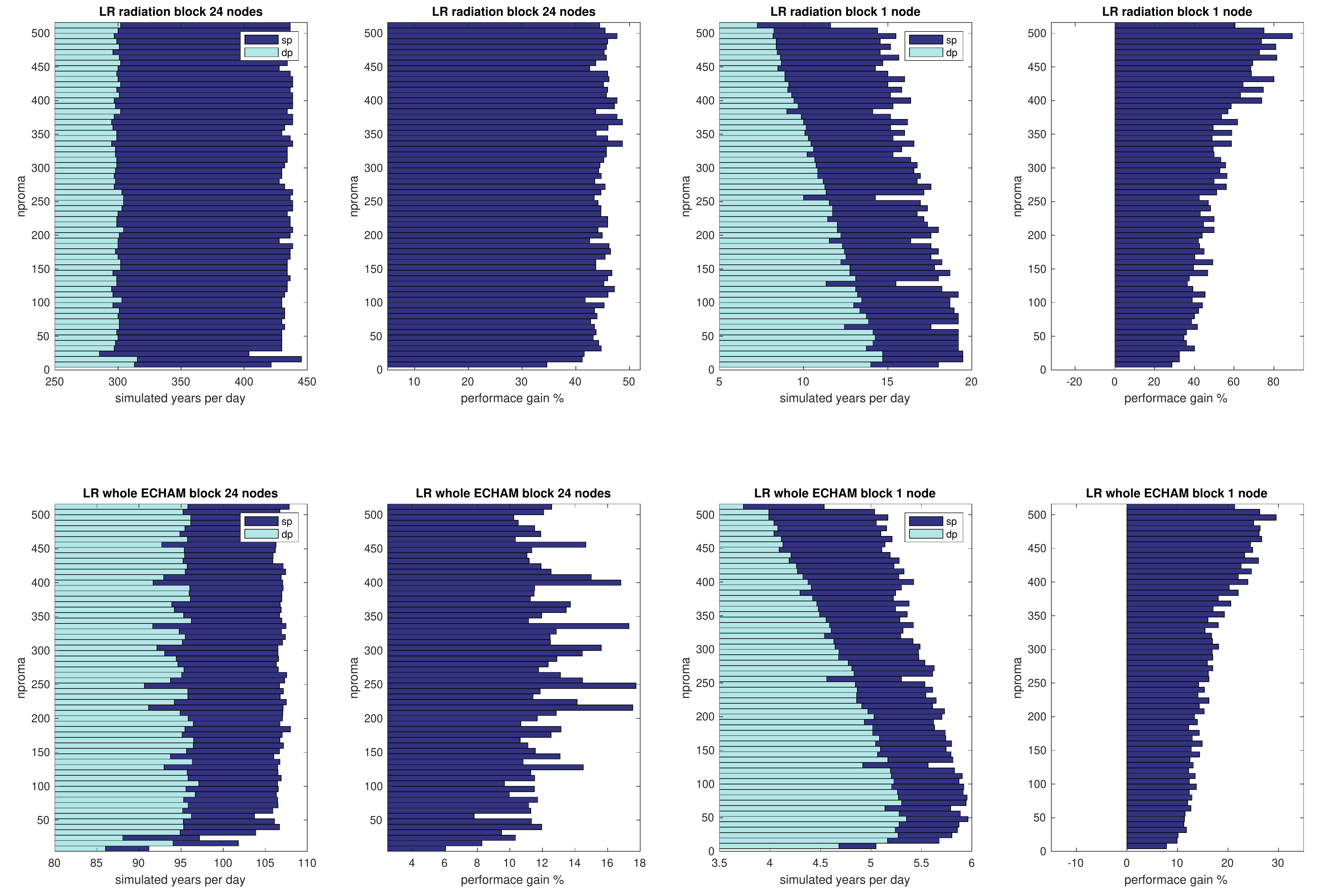}
\caption{As Figure \ref{fig:CR-nproma-block}, but for low resolution (LR) and values of \emph{nproma} between 8 and 512, in steps of 8.
\label{fig:LR-nproma-block}}
\end{figure*}

The difference between the \emph{block} and \emph{cyclic} options are not very significant for all experiments, even though \emph{cyclic}  was slightly faster in some cases. The pictures  for \emph{cyclic} (not presented here) look quite similar. 

Finally we note that measurements for shorter runs of only one month  delivered different optimal values of \emph{nproma}. 
\subsubsection{Best choice of parameter settings for CR configuration}

\begin{figure*}[t]
\includegraphics[width=\textwidth]{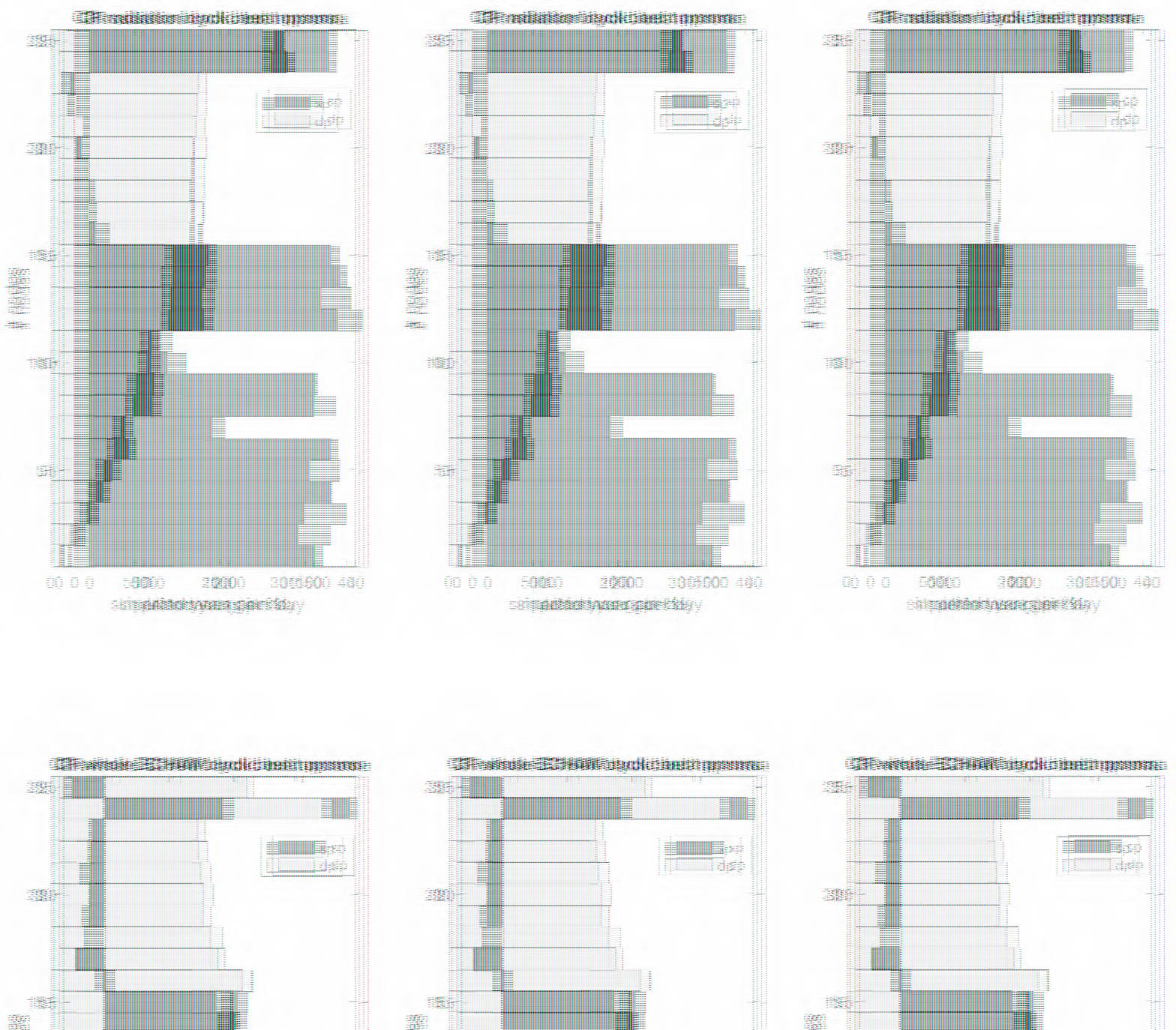}
\caption{As Figure \ref{fig:CR-nproma-block}, but  for 1 to 25 nodes using the respective best value of \emph{nproma}. Cooresponding optimal values can be found in Tables \ref{table:nproma-rad} and \ref{table:nproma-echam}. 
\label{fig:CR-best}}
\end{figure*}

Motivated by the dependency on the parameter \emph{nproma} observed above, we computed the speed-up when using  the fastest choice.  These runs were performed depending on the number of used nodes (from 1 to 25) in the CR configuration, for both \emph{block} and \emph{cyclic} options. The results are shown in Figure \ref{fig:CR-best}. The corresponding best values of \emph{nproma} are given in Tables \ref{table:nproma-rad} and \ref{table:nproma-echam}.

It can be seen that  for an optimal  combination of number of  nodes and \emph{nproma}, the radiation could be accelerated by nearly 40\%. On the other hand, a bad choice of processors (here between 16 and 23)  results in no  performance gain or even a loss.

The speed-up for the whole ECHAM model with \emph{sp} radiation was about  10 to 17\%, when choosing an appropriate combination of nodes and \emph{nproma}.

\begin{table*}[t]
\caption{Best values of parameter \emph{nproma} for different choice of nodes for radiation part.
\label{table:nproma-rad}}
\begin{tabular}{|c|c|c|c|c|c|c|c|c|c|c|c|c|c|c|c|c|c|}
\hline
\# nodes& 1& 2& 3&4&5&6&7&8&9&10&11&12&13&14&15&24&25\\
\hline
\emph{block sp}&16&16 &16 &16 & 76 &16&40& 40& 36  &  84 &   24  &16  &  24&180& 36&228 & 152\\
\emph{block dp} & 16 & 16 & 16 & 16 & 20 & 16&16&56&80&28& 156& 176&32&28 & 88&56& 32\\
\emph{cyclic sp}&48  &  16  &  16  &  24 &   48 &   16    &16  &  96 &   52  & 188 &   28  &  16 &  124& 164&    16 &  252&   212\\
\emph{cyclic dp}&16 &   24  &  16   & 16 &   20  &  16 &   16   &152  & 172 &  136 &  148 &   32  &  40&124   & 16 &   20    &52\\
   \hline
\end{tabular}
\end{table*}

\begin{table*}[t]
\caption{Best values of parameter \emph{nproma} for different choice of nodes for whole ECHAM.
\label{table:nproma-echam}}
\begin{tabular}{|c|c|c|c|c|c|c|c|c|c|c|c|c|c|c|c|c|c|}
\hline
\# nodes& 1& 2& 3&4&5&6&7&8&9&10&11&12&13&14&15&24&25\\
\hline
\emph{block sp}& 48 &48& 32&116&72& 120& 136&44&  36 &152 &120 &56 &24&236 &52 &212& 48\\
\emph{ block dp}&24  &  24 &   32 &   68 &  148 &   16 &   88  & 220 &  84&   232 &   84 &  124 &  252& 36   &228 &  240  & 228\\
\emph{cyclic sp}&48 &   48&    32  & 124   &208    &72  & 192  & 100 &  220  & 152  & 184  & 100  &  92&136  &  68  & 208 &   32\\
\emph{cyclic dp}&24  &   24 &    16  &   24  &  200  &   40  &   40   & 160  &   72  &  128 &   144  &  132 &    28&60   &  84  &   52  &  204\\
   \hline
\end{tabular}
\end{table*}

\subsubsection{Parts of radiation code with highest and lowest speed-up}
We  identified some  subroutines and functions with a very high and some with a very low performance gain by the conversion to sp. They are shown in Tables \ref{tab:fastest} and \ref{tab:slowest}.

A cause that no even higher speed-up was achieved  is that several time-consuming parts (as in  \texttt{rk\_mo\_random\_numbers}) use expensive calculation with integer numbers, taking over 30\% of the total ECHAM time consumption in some cases. Therefore, these parts are not affected by the \emph{sp} conversion.

\begin{table*}[t]
\begin{small}
\caption{Parts of radiation code with highest speed-up by conversion to sp.
\label{tab:fastest} }
\begin{tabular}{|c|c|c|c|c|} 
\hline
Module name&Subroutine/Function&Time \emph{dp} (s) nproma=24&Time \emph{sp} (s) nproma=48&Speed-up (\%)\\
\hline
rk\_mo\_srtm\_solver&delta\_scale\_2d&960.27&413.35&56.95\\
rk\_mo\_echam\_convect\_tables&lookup\_ua\_spline&17.42&7.67&55.97\\
rk\_mo\_rrtm\_coeffs&lrtm\_coeffs&78.59&37.06&52.84\\
rk\_mo\_lrtm\_solver&lrtm\_solver&9815.12&4663.04&52.09\\
rk\_mo\_srtm\_solver&srtm\_solver\_tr&5005.70&2425.09&51.55\\
rk\_mo\_radiation&gas\_profile&27.69&13.57&50.99\\
rk\_mo\_rad\_fastmath&tautrans&3455.26&1790.45&50.53\\
rk\_mo\_rad\_fastmath&transmit&2837.84&1503.41&47.02\\
rk\_mo\_o3clim&o3clim&87.10&47.32&45.67\\
rk\_mo\_aero\_kinne&set\_aop\_kinne&233.65&127.88&45.27\\
\hline
\end{tabular}
\end{small}
\end{table*}

\begin{table*}[t]
\begin{small}
\caption{Parts of radiation code with lowest speed-up by conversion to sp. 
\label{tab:slowest} }
\begin{tabular}{|c|c|c|c|c|} 
\hline
Module name&Subroutine/Function&Time \emph{dp} (s) nproma=24&Time \emph{sp} (s) nproma=48&Speed-up (\%)\\
\hline
rk\_mo\_lrtm\_gas\_optics&gas\_optics\_lw&6517.89&5647.92&13.15\\
rk\_mo\_lrtm\_solver&find\_secdiff&232.42&209.15&10.01\\
rk\_mo\_random\_numbers&m&$1.94\cdot10^{4}$&$1.83\cdot10^{4}$&5.67\\
rk\_mo\_random\_numbers&kissvec&$8.22\cdot10^{4}$&$7.79\cdot10^{4}$&5.23\\
rk\_mo\_lrtm\_driver&planckfunction&2169.69&2070.20&4.59\\
rk\_mo\_srtm\_gas\_optics&gpt\_taumol&4117.74&3931.92&4.51\\
rk\_mo\_random\_numbers&low\_byte&$1.36\cdot10^{4}$&$1.33\cdot10^{4}$&2.20\\\hline
\end{tabular}
\end{small}
\end{table*}

\subsection{Energy Consumption}

We also carried out energy consumption measurements. We used the 
 \emph{IPMI (Intelligent Platform Management Interface)} of the \emph{SLURM} workload manager \emph{ADD} \citep{SchedMD2019}. It is  enabled with the experiment option 
 \begin{quote}
 \begin{verbatim}
 #SBATCH --monitoring=power=5
\end{verbatim}
 \end{quote}
Here we used  one node with the corresponding fast configuration for \emph{nproma} and the option \emph{cyclic}. Simulations were repeated 10 times with a simulation interval of one year. 

As Table \ref{tab:energy} shows, the obtained energy reduction was 13\% and 17\% in blade and cpu power consumption, respectively. We consider these measurements only as a rough estimate. A deeper investigation of energy saving was not the focus of our work.

\begin{table*}[t]
\caption{Energy reduction when using \emph{sp} radiation in ECHAM.
\label{tab:energy}}
\begin{tabular}{|c|c|c|c|}
\hline
Energy Consumption&ECHAM with \emph{dp} radiation (J)&ECHAM with \emph{sp} radiation (J)&Saved energy (\%)\\
\hline
Blade power&803368&698095&13.1\\
CPU power&545762&452408&17.1\\
\hline
\end{tabular}
\end{table*}

\section{Conclusions} 

We have successfully converted the radiation part of ECHAM to single precision arithmetic. All relevant part of the code can now be switched from double to single precision by setting a Fortran \texttt{kind} parameter named \texttt{wp} either to \texttt{dp} or \texttt{sp}. There is one exception where a renaming of subroutines has to be performed. This can be easily done using a (provided) shell script before the compilation of the code. 

We described our incremental conversion process in detail and compared it to other, in this case unsuccessful methods.

We tested the output for the single precision version and found a good agreement with measurement data. The deviations over decadal runs are comparable to the ones of the double precision versions. The difference between the two version lie in the same range.

We achieved an improvement in runtime in coarse and low and resolution of up to 40\% for the radiation itself, and about 10 to 17\% for the whole ECHAM. In this respect, we could support results obtained for the IFS model by \citet{Vana2017}, where the whole model was converted. We also measured an energy saving of about 13 to 17\%. 

 Moreover, we investigated the parts of the code that are sensitive to reduced precision, and those parts which showed comparably high and  low runtime reduction. 
 
 The information we provide may guide other people to convert even more parts of ECHAM to single precision. Moreover, they may also motivate to consider reduced precision arithmetic in other simulation codes.
 
 As a next step, the converted model part will be used in coupled ESM simulation runs over longer time horizons.

\subsection*{Code and data availability}
The code is available in the DKRZ git repository under the URL 
 \url{https://gitlab.dkrz.de/PalMod/echam6-PalMod/~/network/mixed\_precision\_new2} upon request. The conversion scripts (see Appendix \ref{app:script}) and the output data for the single and double precision runs that were used to generate the output plots are available as NetCDF files under \url{https://doi.org/10.5281/zenodo.3560536}.

 \appendix
 \section{Conversion Script}  
  \label{app:script}   
The following shell  script  converts the source code of ECHAM from double  to single precision. 
It renames subroutines and function from the NetCDF library,
 changes a constant in the code to avoid overflow (in a part that was not executed in the used setting), and
sets the constant \texttt{wp} that is used as Fortran \texttt{kind} attribute to the current working precision, either \texttt{dp} or \texttt{sp}.
A script that reverts the changes is analogous.
After the use of one of the two scripts,  the  model has to be re-compiled.
These scripts cannot be used on the standard ECHAM version, but on the one  mentioned in the code availability section.
\begin{quote}
\begin{verbatim}
#!/bin/bash
# script rad_dp_to_sp.sh
# To be executed from the root folder of ECHAM before compilation
for i in ./src/rad_src/rk_mo_netcdf.f90 
         ./src/rad_src/rk_mo_srtm_netcdf.f90 
         ./src/rad_src/rk_mo_read_netcdf77.f90
         ./src/rad_src/rk_mo_o3clim.f90 
         ./src/rad_src/rk_mo_cloud_optics.f90; 
do
   sed -i 's/_double/_real/g' $i  
   sed -i 's/_DOUBLE/_REAL/g' $i
done
sed -i 's/numthresh = 500._wp/numthresh = 75._wp /g' 
       ./src/rad_src/rk_mo_srtm_solver.f90
sed -i 's/INTEGER, PARAMETER :: wp = dp/INTEGER, PARAMETER :: wp = sp/g' 
       ./src/rad_src/rk_mo_kind.f90
echo "ECHAM radiation code converted to single precision."
\end{verbatim}
\end{quote}

\subsection*{Acknowledgements}
This work was supported  by the German Federal Ministry of Education and Research (BMBF) as a Research for Sustainability initiative (FONA) through the project PalMod (FKZ: 01LP1515B), Working Group 4, Work Package 4.4.
The authors  wish to thank 
Mohammad Reza Heidari, Hendryk Bockelmann and J{\"o}rg Behrens from the German Climate Computing Center DKRZ in Hamburg, Uwe Mikolajewicz and Sebastian Rast from the Max-Planck Institute for Meteorology in Hamburg, Peter D{\"u}ben from the ECMWF in Reading, Robert Pincus from Colorado University, Zhaoyang Song from the Helmholtz Centre for Ocean Research Kiel GEOMAR and Gerrit Lohmann from AWI Bremerhaven and Bremen University
for their valuable help and suggestions.

\bibliographystyle{alpha}
\bibliography{ms}

\end{document}